\input harvmac.tex
\input epsf
\noblackbox

%%% Paragraphs
\newcount\figno

\figno=0
\def\fig#1#2#3{
\par\begingroup\parindent=0pt\leftskip=1cm\rightskip=1cm\parindent=0pt
\baselineskip=11pt \global\advance\figno by 1 \midinsert
\epsfxsize=#3 \centerline{\epsfbox{#2}} \vskip 12pt
\centerline{{\bf Figure \the\figno :}{\it ~~ #1}}\par
\endinsert\endgroup\par}
\def\figlabel#1{\xdef#1{\the\figno}}

%%% special math symbols
\font\cmss=cmss10
\font\cmsss=cmss10 at 7pt
\def\half{{1\over 2}}
\def\rlx{\relax\leavevmode}
\def\inbar{\vrule height1.5ex width.4pt depth0pt}
\def\IC{\relax\,\hbox{$\inbar\kern-.3em{\rm C}$}}
\def\IR{\relax{\rm I\kern-.18em R}}
\def\IN{\relax{\rm I\kern-.18em N}}
\def\IP{\relax{\rm I\kern-.18em P}}
\def\ZZ{\rlx\leavevmode\ifmmode\mathchoice{\hbox{\cmss Z\kern-.4em Z}}
 {\hbox{\cmss Z\kern-.4em Z}}{\lower.9pt\hbox{\cmsss Z\kern-.36em Z}}
 {\lower1.2pt\hbox{\cmsss Z\kern-.36em Z}}\else{\cmss Z\kern-.4em Z}\fi}

%%% misc.
\def\narrowplus{\kern -.04truein + \kern -.03truein}
\def\narrowminus{- \kern -.04truein}
\def\narrowminussub{\kern -.02truein - \kern -.01truein}

\def\Im{{\rm Im}}
\def\Re{{\rm Re}}
\def\F{{\cal F}}
\def\N{{\cal N}}
\def\T{{\cal T}}
\def\I{{\cal I}}
\def\R{{\cal R}}

\def\mte{m\!\times\! e}

\def\o#1{\overline{#1}}

\def\cK{{\cal{K}}}
\def\R{{\cal{R}}}
\def\F{{\cal{F}}}
\def\N{{\cal N}}

\def\I{{\cal I}}

%%% further macros

%%% References

%\AngelantonjCT
\lref\rasreview{
C.~Angelantonj and A.~Sagnotti,
{\it Open Strings},
hep-th/0204089.
%%CITATION = HEP-TH 0204089;%%
}

%\AngelantonjFT
\lref\raaads{
C.~Angelantonj, I. Antoniadis, G. D'Appollonio, E. Dudas and A.~Sagnotti,
{\it Type I vacua with brane supersymmetry breaking},
 Nucl.\ Phys. {\bf B572} (2000) 36, hep-th/9911081.
%%CITATION = HEP-TH 9911081;%%
}

%\DoranVE
\lref\DoranVE{
C.~F.~Doran, M.~Faux and B.~A.~Ovrut,
{\it Four-dimensional N = 1 Super Yang-Mills Theory from an M
theory Orbifold},
hep-th/0108078.
%%CITATION = HEP-TH 0108078;%%
}

%\FauxSP
\lref\FauxSP{
M.~Faux, D.~L\"ust and B.~A.~Ovrut,
{\it An M-theory Perspective on Heterotic K3 Orbifold Compactifications},
hep-th/0010087.
%%CITATION = HEP-TH 0010087;%%
}

%\FauxDV
\lref\FauxDV{
M.~Faux, D.~L\"ust and B.~A.~Ovrut,
{\it Local Anomaly Cancellation, M-theory Orbifold and Phase-transitions},
Nucl.\ Phys.\ B {\bf 589} (2000) 269,
hep-th/0005251.
%%CITATION = HEP-TH 0005251;%%
}

%\FauxHM
\lref\FauxHM{
M.~Faux, D.~L\"ust and B.~A.~Ovrut,
{\it Intersecting Orbifold Planes and Local
Anomaly Cancellation in  M-theory},
Nucl.\ Phys.\ B {\bf 554} (1999) 437,
hep-th/9903028.
%%CITATION = HEP-TH 9903028;%%
}

\lref\rfaux{C.F.~Doran, M.~Faux, {\it
Intersecting Branes in M-Theory and Chiral Matter in Four Dimensions},
JHEP {\bf 0208} (2002) 024, hep-th/0207162.
%%CITATION = HEP-TH 0207162 ;%%
}

\lref\rpradisi{G.~Pradisi, {\it Magnetized (Shift-)Orientifolds}, hep-th/0210088.
%%CITATION = HEP-TH 0210088 ;%%
}

\lref\rkokora{C.~Kokorelis, {\it Deformed Intersecting D6-Brane GUTS I},
hep-th/0209202.
%%CITATION = HEP-TH  0209202;%%
}

\lref\rkokorb{C.~Kokorelis, {\it Deformed Intersecting D6-Brane GUTS II},
hep-th/0210200.
%%CITATION = HEP-TH 0210200 ;%%
}

%\Mari\~noAF
\lref\MarinoAF{ M.~Mari\~no, R.~Minasian, G.~W.~Moore and
A.~Strominger, {\it Nonlinear Instantons from Supersymmetric
$p$-branes}, JHEP {\bf 0001} (2000) 005, hep-th/9911206.
%%CITATION = HEP-TH 9911206;%%
}

\lref\rhanany{A. Hanany and B. Kol, {\it
On Orientifolds, Discrete Torsion, Branes and M Theory},
JHEP {\bf 0006} (2000) 013,  hep-th/0003025.
%%CITATION = HEP-TH  0003025;%%
}

\lref\rangles{M.~Berkooz, M.~R.~Douglas and R.~G.~Leigh, {\it Branes Intersecting
at Angles}, Nucl. Phys. B {\bf 480} (1996) 265, hep-th/9606139.
%%CITATION = HEP-TH 9606139;%%
}

%\KleinVU
\lref\KleinVU{ M.~Klein, {\it Couplings in Pseudo-Supersymmetry},
Phys.\ Rev. {\bf D66} (2002) 055009,hep-th/0205300.
%%CITATION = HEP-TH 0205300;%%
}

\lref\rKleinb{C.P. Burgess, E. Filotas, M. Klein, F. Quevedo,
{\it  Low-Energy Brane-World Effective Actions and Partial Supersymmetry Breaking},
hep-th/0209190.
%%CITATION = HEP-TH 0209190;%%
}

\lref\rbailinb{D.~Bailin, G.V.~Kraniotis and A.~Love, {\it
Standard-like models from intersecting D5-branes},  hep-th/0210227.
%%CITATION = HEP-TH 0210227;%%
}

\lref\rbailina{D.~Bailin, G.V.~Kraniotis and A.~Love, {\it
New Standard-like Models from Intersecting D4-Branes},  hep-th/0208103.
%%CITATION = HEP-TH 0208103;%%
}

\lref\rrab{R.~Rabadan, {\it Branes at Angles, Torons, Stability and
Supersymmetry}, Nucl.\ Phys.\ B {\bf 620} (2002) 152, hep-th/0107036.
%%CITATION = HEP-TH 0107036;%%
}

\lref\rbgkb{R.~Blumenhagen, L.~G\"orlich and B.~K\"ors, {\it
Supersymmetric Orientifolds in 6D with D-Branes at Angles}, Nucl.\
Phys. {\bf B569} (2000) 209, hep-th/9908130.
%%CITATION = HEP-TH 9908130;%%
}

\lref\rbgo{R.~Blumenhagen, L.~G\"orlich and T.~Ott,
{\it Supersymmetric Intersecting Branes on the Type IIA $T^6/Z_4$ orientifold},
JHEP {\bf 0301} (2003) 021, hep-th/0211059.
%%CITATION = HEP-TH 0211059;%%
}

\lref\rbgkc{R.~Blumenhagen, L.~G\"orlich and B.~K\"ors, {\it
Supersymmetric 4D Orientifolds of Type IIA with D6-branes at Angles},
JHEP {\bf 0001} (2000) 040, hep-th/9912204.
%%CITATION = HEP-TH 9912204;%%
}

\lref\rfhs{S.~F\"orste, G.~Honecker and R.~Schreyer, {\it Supersymmetric
$\ZZ_N \times \ZZ_M$ Orientifolds in 4-D with D-branes at Angles},
Nucl. Phys. B {\bf 593} (2001) 127, hep-th/0008250.
%%CITATION = HEP-TH 0008250;%%
}

\lref\rfhstwo{S.~F\"orste, G.~Honecker and R.~Schreyer, {\it Orientifolds
with Branes at Angles}, JHEP {\bf 0106} (2001) 004, hep-th/0105208.
%%CITATION = HEP-TH 0105208;%%
}

\lref\rbgklnon{R.~Blumenhagen, L.~G\"orlich, B.~K\"ors and D.~L\"ust,
{\it Noncommutative Compactifications of Type I Strings on Tori with Magnetic
Background Flux}, JHEP {\bf 0010} (2000) 006, hep-th/0007024.
%%CITATION = HEP-TH 0007024;%%
}

\lref\rangela{R.~Blumenhagen and C.~Angelantonj,
{\it Discrete Deformations in Type I Vacua},
Phys.Lett. {\bf B473} (2000) 86, hep-th/9911190.
%%CITATION = HEP-TH 9911190;%%
}

\lref\rmayr{P. Mayr, {\it On Supersymmetry Breaking in String Theory and 
its Realization in Brane Worlds}, Nucl.Phys. {\bf B593} (2001) 99, 
hep-th/0003198.
%%CITATION = HEP-PH 0003198;%%
}

\lref\ritaa{R. D'Auria, S. Ferrara and S. Vaula, {\it
N=4 gauged supergravity and a IIB orientifold with fluxes},
New J.Phys. {\bf 4} (2002) 71, hep-th/0206241.
%%CITATION = HEP-PH 0206241;%%
}

\lref\ritab{L. Andrianopoli, R. D'Auria, S. Ferrara and  M. A. Lledo, 
{\it 4-D gauged supergravity analysis of Type IIB vacua on 
$K3\times T^2/Z_2$}, hep-th/0302174.
%%CITATION = HEP-PH 0302174;%%
}

%\Iba\~nezDJ
\lref\IbanezDJ{ L.~E.~Iba\~nez, {\it Standard Model Engineering
with Intersecting Branes}, hep-ph/0109082.
%%CITATION = HEP-PH 0109082;%%
}

\lref\rbgklmag{R.~Blumenhagen, L.~G\"orlich, B.~K\"ors and
D.~L\"ust, {\it Magnetic Flux in Toroidal Type I
Compactification}, Fortsch. Phys. {\bf 49} (2001) 591,
hep-th/0010198.
%%CITATION = HEP-TH 0010198;%%
}

\lref\rba{C.~Angelantonj and R.~Blumenhagen, {\it Discrete
Deformations in Type I Vacua}, Phys. Lett. B {\bf 473} (2000) 86,
hep-th/9911190.
%%CITATION = HEP-TH 9911190;%%
}

\lref\ras{C.~Angelantonj and A.~Sagnotti, {\it Type I Vacua and
Brane Transmutation}, hep-th/0010279.
%%CITATION = HEP-TH 0001279;%%
}

\lref\raads{C.~Angelantonj, I.~Antoniadis, E.~Dudas and
A.~Sagnotti, {\it Type I Strings on Magnetized Orbifolds and Brane
Transmutation}, Phys. Lett. B {\bf 489} (2000) 223,
hep-th/0007090.
%%CITATION = HEP-TH 0007090;%%
}

\lref\rbkl{R.~Blumenhagen, B.~K\"ors and D.~L\"ust,
{\it Type I Strings with $F$ and $B$-Flux}, JHEP {\bf 0102} (2001) 030,
hep-th/0012156.
%%CITATION = HEP-TH 0012156;%%
}

\lref\rbgkl{R.~Blumenhagen, L.~G\"orlich, B.~K\"ors and D.~L\"ust,
{\it Asymmetric Orbifolds, Noncommutative Geometry and Type I
Vacua}, Nucl.\ Phys.\ B {\bf 582} (2000) 44, hep-th/0003024.
%%CITATION = HEP-TH 0003024;%%
}

\lref\rbbkl{R.~Blumenhagen, V.~Braun, B.~K\"ors and D.~L\"ust,
{\it Orientifolds of K3 and Calabi-Yau Manifolds with Intersecting D-branes},
JHEP {\bf 0207} (2002) 026, hep-th/0206038.
%%CITATION = HEP-TH 0206038;%%
}

\lref\rbbklb{R.~Blumenhagen, V.~Braun, B.~K\"ors and D.~L\"ust,
{\it The Standard Model on the Quintic}, hep-th/0210083.
%%CITATION = HEP-TH 0210083;%%
}

\lref\rura{A.M.~Uranga,
{\it Local models for intersecting brane worlds}, hep-th/0208014.
%%CITATION = HEP-TH 0208014;%%
}

\lref\rcvetica{M.~Cvetic, G.~Shiu and  A.~M.~Uranga,  {\it
Three-Family Supersymmetric Standard-like Models from Intersecting
Brane Worlds}, Phys. Rev. Lett. {\bf 87} (2001) 201801,
hep-th/0107143.
%%CITATION = HEP-TH 0107143;%%
}

\lref\rcveticb{M.~Cvetic, G.~Shiu and  A.~M.~Uranga,  {\it
Chiral Four-Dimensional N=1 Supersymmetric Type IIA Orientifolds from
Intersecting D6-Branes}, Nucl. Phys. B {\bf 615} (2001) 3, hep-th/0107166.
%%CITATION = HEP-TH 0107166;%%
}

\lref\rott{R.~Blumenhagen, B.~K\"ors, D.~L\"ust and T.~Ott, {\it
The Standard Model from Stable Intersecting Brane World Orbifolds},
Nucl. Phys. B {\bf 616} (2001) 3, hep-th/0107138.
%%CITATION = HEP-TH 0107138;%%
}

\lref\rottb{R.~Blumenhagen, B.~K\"ors, D.~L\"ust and T.~Ott, {\it
Intersecting Brane Worlds on Tori and Orbifolds}, hep-th/0112015.
%%CITATION = HEP-TH 0112015;%%
}

\lref\rbonna{S.~F\"orste, G.~Honecker and R.~Schreyer, {\it
Orientifolds with Branes at Angles}, JHEP {\bf 0106} (2001) 004,
hep-th/0105208.
%%CITATION = HEP-TH 0105208;%%
}

\lref\rbonnb{G.~Honecker, {\it Intersecting Brane World Models from
D8-branes on $(T^2 \times T^4/\ZZ_3)/\Omega R_1$ Type IIA Orientifolds},
JHEP {\bf 0201} (2002) 025, hep-th/0201037.
%%CITATION = HEP-TH 0201037;%%
}

\lref\rqsusy{D.~Cremades, L.~E.~Iba\~nez and F.~Marchesano, {\it
SUSY Quivers, Intersecting Branes and the Modest Hierarchy
Problem}, JHEP {\bf 0207} (2002) 009, hep-th/0201205.
%%CITATION = HEP-TH 0201205;%%
}

\lref\rqsusyb{D.~Cremades, L.~E.~Iba\~nez and F.~Marchesano, {\it
     Intersecting Brane Models of Particle Physics and the Higgs Mechanism},
   JHEP {\bf 0207} (2002) 022,   hep-th/0203160.
%%CITATION = HEP-TH 0203160;%%
}

%\AldazabalPY
\lref\AldazabalPY{ G.~Aldazabal, L.~E.~Iba\~nez and A.~M.~Uranga,
{\it Gauging Away the Strong CP Problem}, hep-ph/0205250.
%%CITATION = HEP-PH 0205250;%%
}

\lref\hw{P.S. Howe and P.C. West, {\it The Complete N=2, D=10 Supergravity},
Nucl.\ Phys.\ B 238 (1984) 181.}

\lref\gp{M. Grana and J. Polchinski,
{\it Gauge/Gravity Duals with Holomorphic
  Dilaton}, Phys.\ Rev. {\bf D65} (2002) 126005, hep-th/0106014.
%%CITATION = HEP-TH 0106014;%%
}

\lref\bachas{C.~Bachas, {\it A Way to Break Supersymmetry},
hep-th/9503030.
%%CITATION = HEP-TH 9503030;%%
}

\lref\rafiruph{G.~Aldazabal, S.~Franco, L.~E.~Iba\~nez, R.~Rabadan
and A.~M.~Uranga, {\it Intersecting Brane Worlds}, JHEP {\bf 0102}
(2001) 047, hep-ph/0011132.
%%CITATION = HEP-PH 0011132;%%
}

\lref\rafiru{G.~Aldazabal, S.~Franco, L.~E.~Iba\~nez, R.~Rabadan
and A.~M.~Uranga, {\it $D=4$ Chiral String Compactifications from
Intersecting Branes}, J.\ Math.\ Phys.\  {\bf 42} (2001) 3103,
hep-th/0011073.
%%CITATION = HEP-TH 0011073;%%
}

\lref\rimr{L.~E.~Iba\~nez, F.~Marchesano and R.~Rabadan, {\it
Getting just the Standard Model at Intersecting Branes}, JHEP {\bf
0111} (2001) 002, hep-th/0105155.
%%CITATION = HEP-TH 0105155;%%
}

\lref\belrab{J.~Garcia-Bellido and R.~Rabadan, {\it Complex Structure Moduli
Stability in Toroidal Compactifications},
JHEP {\bf 0205} (2002) 042, hep-th/0203247.
%%CITATION = HEP-TH 0203247;%%
}

\lref\rcim{D.~Cremades, L.~E.~Iba\~nez and F.~Marchesano, {\it
    Standard Model at Intersecting D5-branes: Lowering the String Scale},
   Nucl.\ Phys. {\bf B643} (2002) 93,  hep-th/0205074.
%%CITATION = HEP-TH 0205074;%%
}

\lref\rkokoa{C.~Kokorelis, {\it GUT Model Hierarchies from Intersecting Branes},
JHEP {\bf 0208} (2002) 018, hep-th/0203187.
%%CITATION = HEP-TH 0203187;%%
}

\lref\rkokob{C.~Kokorelis, {\it New Standard Model Vacua from Intersecting Branes},
JHEP {\bf 0209} (2002) 029, hep-th/0205147.
%%CITATION = HEP-TH 0205147;%%
}

%\HoneckerDJ
\lref\HoneckerDJ{
G.~Honecker,
{\it Non-supersymmetric Orientifolds with D-branes at Angles}, hep-th/0112174.
%%CITATION = HEP-TH 0112174;%%
}

\lref\rcls{M.~Cvetic, P.~Langacker, and G.~Shiu, {\it
 Phenomenology of A Three-Family Standard-like String Model},
Phys.\ Rev. {\bf D66} (2002) 066004, hep-ph/0205252.
%%CITATION = HEP-TH 0205252;%%
}

\lref\rclsb{M.~Cvetic, P.~Langacker, and G.~Shiu, {\it
 A Three-Family Standard-like Orientifold Model: Yukawa Couplings and Hierarchy},
Nucl.\ Phys. {\bf B642} (2002) 139, hep-ph/0206115.
%%CITATION = HEP-TH 0206115;%%
}

\lref\rgkp{S.~B.~Giddings, S.~Kachru and J.~Polchinski, {\it
Hierarchies from Fluxes in String Compactifications},
hep-th/0105097.
%%CITATION = HEP-TH 0105097;%%
}

%\DasguptaSS
\lref\DasguptaSS{
K.~Dasgupta, G.~Rajesh and S.~Sethi,
{\it M theory, orientifolds and G-flux},
JHEP {\bf 9908}, 023 (1999), hep-th/9908088.
%%CITATION = HEP-TH 9908088;%%
}

\lref\rkst{S.~Kachru, M.~Schulz and S.~Trivedi, {\it
             Moduli Stabilization from Fluxes in a Simple IIB Orientifold},
              hep-th/0201028.
%%CITATION = HEP-TH 0201028;%%
}

\lref\rkstt{S.~Kachru, M.~Schulz, P.K. Tripathy and S.~Trivedi, {\it
New Supersymmetric String Compactifications}, 
 hep-th/0211182.
%%CITATION = HEP-TH 0211182;%%
}

\lref\ruranga{A.~M.~Uranga, {\it D-brane, Fluxes and Chirality},
hep-th/0201221.
%%CITATION = HEP-TH 0201221;%%
}

\lref\ruram{A.~M.~Uranga, {\it Localized Instabilities at Conifolds},
hep-th/0204079.
%%CITATION = HEP-TH 0204079;%%
}

\lref\rpark{M. Mihailescu, I.Y. Park and  T.A. Tran, {\it
D-branes as Solitons of an N=1, D=10 Non-commutative Gauge Theory},
Phys.Rev. {\bf D64} (2001) 046006, hep-th/0011079.
%%CITATION = HEP-TH 0011079;%%
}

%\BailinIE
\lref\BailinIE{
D.~Bailin, G.~V.~Kraniotis and A.~Love,
{\it Standard-like Models from Intersecting D4-branes},
Phys.\ Lett.\ B {\bf 530} (2002) 202,
hep-th/0108131.
%%CITATION = HEP-TH 0108131;%%
}

\lref\rangi{C.~Angelantonj, I.~Antoniadis, G.~D'Appollonio, E.~Dudas
     and A.~Sagnotti, {\it Type I vacua with brane supersymmetry breaking},
Nucl.\ Phys. {\bf B572} (2000) 36, hep-th/9911081.
%%CITATION = HEP-TH 9911081;%%
}

\lref\rwitti{E.~Witten, {\it
            BPS Bound States of D0-D6 and D0-D8 Systems in a B-Field},
               JHEP {\bf 0204} (2002) 012,  hep-th/0012054.
%%CITATION = HEP-TH 0012054;%%
}

\lref\tv{T.R.\ Taylor and C.\ Vafa,
{\it RR flux on Calabi-Yau and partial supersymmetry breaking},
Phys.\ Lett.\ B {\bf 474}, 130 (2000), hep-th/9912152.}

\lref\pols{J.\ Polchinski and A.\ Strominger,
{\it New Vacua for Type II String Theory},
Phys.\ Lett.\ B {\bf 388} (1996) 736, hep-th/9510227.}

\lref\agnt{I.\ Antoniadis, E. Gava, K.S.\ Narain and T.R.\ Taylor,
{\it Duality in superstring compactifications with magnetic field
backgrounds}, Nucl.\ Phys.\ B {\bf 511} (1998) 611,
hep-th/9708075.}

\lref\apt{I.\ Antoniadis, H.\ Partouche and T.R.\ Taylor,
{\it Spontaneous Breaking of N=2 Global Supersymmetry},
Phys.\ Lett.\ B {\bf 372} (1996) 83, hep-th/9512006.}

\lref\ferrara{S.~Ferrara and M.~Porrati, {\it N = 1 no-scale
supergravity from IIB orientifolds}, Phys.\ Lett.\ B {\bf 545}
(2002) 411, hep-th/0207135; L.~Andrianopoli, R.~D'Auria,
S.~Ferrara and M.~A.~Lledo, {\it Gauged extended supergravity
without cosmological constant: No-scale  structure and
supersymmetry breaking}, hep-th/0212141; L.~Andrianopoli,
R.~D'Auria, S.~Ferrara and M.~A.~Lledo, {\it N = 2 super-Higgs, N
= 1 Poincar\'e vacua and quaternionic geometry}, JHEP {\bf 0301}
(2003) 045, hep-th/0212236; L.~Andrianopoli, R.~D'Auria,
S.~Ferrara and M.~A.~Lledo, {\it 4-D supergravity analysis of Type
IIB vacua on $K_3\times T^2/Z_2$}, hep-th/0302174.}

\lref\as{J.J. Atick and A. Sen, {\it Covariant one Loop Fermion
Emission Amplitudes in Closed String Theories},
Nucl.\ Phys.\ {\bf B293} (1987) 317.}

\lref\caf{A. Ceresole, R. D'Auria and S. Ferrara, {\it The
Symplectic Structure of N=2 Supergravity and its Central
Extension}, Nucl. Phys. Proc. Suppl. {\bf 46} (1996) 67,
hep-th/9509160.
%%CITATION = HEP-TH 9509160;%%
}

\lref\curioa{G. Curio, A. Klemm, D. L\"ust and S. Theisen, {\it
On the Vacuum Structure of Type II String Compactifications on Calabi-Yau
Spaces with H-Fluxes}, Nucl.\ Phys. {\bf B609} (2001) 3, hep-th/0012213.
%%CITATION = HEP-TH 0012213;%%
}

\lref\curiob{G. Curio, A. Klemm, B. K\"ors and D. L\"ust, {\it
 Fluxes in Heterotic and Type II String Compactifications},
Nucl.\ Phys. {\bf B620} (2002) 237, hep-th/0106155.
%%CITATION = HEP-TH 0106155;%%
}

\lref\curioc{G. Curio, B. K\"ors and D. L\"ust, {\it
Fluxes and Branes in Type II Vacua and M-theory Geometry with G(2) and Spin(7)
Holonomy}, Nucl.\ Phys. {\bf B636} (2002) 197, hep-th/0111165.
%%CITATION = HEP-TH 0111165;%%
}

\lref\micu{S. Gurrieri, J. Louis, A. Micu and D. Waldram, {\it
Mirror Symmetry in Generalized Calabi-Yau Compactifications},
hep-th/0211102.
%%CITATION = HEP-TH 0211102;%%
}

%\BlumenhagenUA
\lref\BlumenhagenUA{
R.~Blumenhagen, B.~K\"ors, D.~L\"ust and T.~Ott,
{\it Hybrid Inflation in Intersecting Brane Worlds},
Nucl.\ Phys. {\bf B641} (2002) 235, hep-th/0202124.
%%CITATION = HEP-TH 0202124;%%
}

%\AganagicGS
\lref\AganagicGS{
M.~Aganagic and C.~Vafa,
{\it Mirror Symmetry, D-branes and Counting Holomorphic Discs}
hep-th/0012041.
%%CITATION = HEP-TH 0012041;%%
}

%\Pradisi
\lref\refPradisi{
G.~Pradisi,
{\it Type I Vacua from Diagonal $\ZZ_3$-Orbifolds},
Nucl.\ Phys.\ B {\bf 575} (2000) 134,
hep-th/9912218.
%%CITATION = HEP-TH 9912218;%%
}

\lref\rbbhl{K. Becker, M. Becker, M. Haack and  J. Louis, {\it
Supersymmetry Breaking and alpha'-Corrections to Flux Induced Potentials},
JHEP {\bf 0206} (2002) 060,  hep-th/0204254.
%%CITATION = HEP-TH 0204254;%%
}

\lref\rwolf{O. DeWolfe and  S.B. Giddings, {\it
Scales and hierarchies in warped compactifications and brane worlds},
 hep-th/0208123.
%%CITATION = HEP-TH 0208123;%%
}

\lref\rfreya{A.R. Frey and  J. Polchinski, {\it
 N=3 Warped Compactifications},
Phys.\ Rev. {\bf D65} (2002) 126009, hep-th/0201029.
%%CITATION = HEP-TH 0201029;%%
}

\lref\rfreyb{A.R. Frey and A. Mazumdar, {\it
3-Form Induced Potentials, Dilaton Stabilization, and Running Moduli},
hep-th/0210254.
%%CITATION = HEP-TH 0210254;%%
}

\lref\rkklt{S. Kachru, R. Kallosh, A. Linde and S.P. Trivedi, {\it
de Sitter Vacua in String Theory},
hep-th/0301240.
%%CITATION = HEP-TH 0301240;%%
}

\lref\rtt{P.K. Tripathy and S.P. Trivedi,  {\it
Compactification with Flux on K3 and Tori},
hep-th/0301139.
%%CITATION = HEP-TH 0301139;%%
}

\lref\rgvw{S. Gukov, C. Vafa and E. Witten, {\it CFT's From
Calabi-Yau Four-folds}, Nucl.\ Phys. {\bf B584} (2000) 69,
Erratum-ibid. {\bf B608} (2001) 477, hep-th/9906070.
%%CITATION = HEP-TH 9906070;%%
}

\lref\rgukov{S. Gukov, {\it
Solitons, superpotentials and calibrations},
Nucl. Phys. {\bf B574} (2000) 169,  hep-th/9911011.
%%CITATION = HEP-TH 9911011;%%
}

\lref\lustie{D. L\"ust and S. Stieberger, {\it Gauge Threshold Corrections
in Intersecting Brane World Models},
hep-th/0302221.
%%CITATION = HEP-TH 0302221;%%
}

%\CremadesQJ
\lref\CremadesQJ{
D.~Cremades, L.~E.~Ibanez and F.~Marchesano,
{\it Yukawa couplings in intersecting D-brane models},
hep-th/0302105.
%%CITATION = HEP-TH 0302105;%%
}

%%% Title page
\Title{\vbox{
 \hbox{HU--EP-03/06}
 \hbox{CERN--TH/2003-032}
 \hbox{hep-th/0303016}}}
 %\vskip-1cm
{\vbox{\centerline{Moduli Stabilization in  Chiral Type IIB}
\vskip 0.3cm \centerline{Orientifold Models with Fluxes} }}
\centerline{Ralph Blumenhagen{$^1$}, Dieter L\"ust{$^1$}, Tomasz
R.\ Taylor{$^{2,\star}$ }}
\bigskip\medskip
\centerline{$^1${\it Humboldt-Universit\"at zu Berlin, Institut f\"ur
Physik,}}
\centerline{\it Invalidenstrasse 110, 10115 Berlin, Germany}
\centerline{\tt e-mail:
blumenha, luest@physik.hu-berlin.de}
\centerline{$^2${\it CERN Theory Division, CH-1211 Geneva 23, Switzerland}}
\centerline{\tt e-mail: taylor@neu.edu}
\bigskip
\bigskip

\centerline{\bf Abstract}
\noindent
We consider Type IIB orientifold models on Calabi-Yau spaces
with three-form G-flux turned on. These fluxes freeze
some of the complex structure moduli and the complex dilaton
via an F-term scalar potential. By introducing pairs of
D9-$\o{\rm D9}$ branes
with abelian magnetic fluxes it is possible to freeze also
some of the K\"ahler moduli via a D-term potential. Moreover,
such magnetic fluxes in general lead to chiral fermions, which
make them interesting for string model-building.
These issues are demonstrated  in a  simple toy model based on
a $\ZZ_2\times \ZZ_2'$ orbifold.

\vfill\hrule\vskip 1mm\noindent {$^{\star}$\ninerm On sabbatical
leave from Department of Physics, Northeastern University, Boston,
MA 02115, USA.} \Date{}
%%% text
\newsec{Introduction}

There exist many obstacles that string theory has to overcome in
order to make  contact with low-energy physics. The list of
problems contains in particular the question of how to remove the
huge vacuum degeneracy, i.e.\ how to fix the moduli that are
typically present in any string compactification. Second, the
light string modes must reproduce the spectrum of the standard
model; hence realistic string vacua must naturally lead to chiral
fermions. A third, very important issue is the problem of
space-time supersymmetry breaking, which eventually has to be
achieved without creating any new vacuum instabilities. With the
advent of D-branes and also with the introduction of background
fluxes in the internal space, some new perspectives in addressing
these questions arose during the last few years.

First, in the context of intersecting brane world models
\refs{\rangles
\rbgklnon\raads\rbgklmag\ras\rafiru\rafiruph\rbkl\rimr\rbonna\rrab
\rott\rcvetica\rcveticb\BailinIE\rqsusy
\BlumenhagenUA\rqsusyb\rkokoa\rbbkl\rura\rbbklb\rpradisi\rbgo\CremadesQJ-\lustie} 
with type IIA
D6-branes wrapped around 3-cycles of the internal Calabi-Yau
space, it is possible to construct in a systematic way orientifold
compactifications with standard model-like spectra with chiral
fermions. Part of the space-time supersymmetry is preserved if the
intersecting D6-branes wrap supersymmetric (special lagrangian)
3-cycles, which must be calibrated with respect to the same
holomorphic 3-form as the O6-planes are. In general, the tension
of the D6-branes and of the O6-planes introduces a vacuum energy,
which is described by a Fayet-Iliopolous term in the language of
$N=1$ supersymmetric field theory \rqsusy. These D-terms depend
only on (part of) the complex structure moduli, which can be fixed
upon minimization of the potential. In the (T-dual) Type IIB
mirror picture, one is dealing with magnetic gauge fluxes on the
world-volume of D9-branes \refs{\rbgklnon,\rbbkl}. Since the Type
IIB F-flux
%which is in one to one
%correspondence to the IIA D6-brane
%intersection angles,
is integrated over 2-cycles of the Calabi-Yau space,
the D-term potential now stabilizes (part of) the Type IIB K\"ahler moduli.

Another recent approach to moduli stabilization involves Type IIB
background 3-form fluxes on the internal Calabi-Yau manifold,
$G=\tau\, H_3+F_3$, where  $H_3$ originates from the NS-NS sector,
$F_3$ from the R-R sector and $\tau$ is the dilaton that
``complexifies'' the flux \pols. These so-called G-fluxes give
rise to a scalar potential, which freezes many of the complex
structure moduli of the Calabi-Yau and the dilaton. In addition,
supersymmetry may get partially or completely broken. A very
convenient way of analyzing the consequences of turning on G-fluxes
is to use an effective superpotential that can be computed from
the ten-dimensional kinetic terms for the 3-forms \tv. Using
the effective F-term (super)potential, the vacuum structure of
this type of flux compactifications was recently discussed in
several papers
\refs{\DasguptaSS\rmayr\curioa\rgkp\curiob\curioc\rkst\rfreya
\rbbhl\ritaa\rwolf\rfreyb\micu\rkstt\rtt-\rkklt}. Upon inspection of the
induced potential, one realizes that turning on these fluxes on a
compact Calabi-Yau space in a local way, i.e. such that $\int
H_3\wedge F_3=0$, the minima of the potential are generically at
those points in the Calabi-Yau moduli space where the geometry
degenerates and supersymmetry is restored \tv. So far, partial
supersymmetry breaking was only shown to be possible at certain
(conifold) points of a {\it non-compact} Calabi-Yau space. As we
will discuss in this paper, by choosing non-local fluxes with
 $\int H_3\wedge F_3\neq 0$, partial supersymmetry breaking and moduli
stabilization can  also be achieved at points where the {\it compact}
Calabi-Yau space is non-degenerate.

The main aim of this paper is to combine the G-flux
compactifications with the scenario of Type IIA intersecting
branes or, respectively, with Type IIB magnetic gauge  fluxes on
D9-branes. For concreteness, we will apply the general formalism
to a Type IIB $\ZZ_2\otimes \ZZ_2'$ orientifold model with O3-
and three sets of O7-planes. We turn on both non-trivial G-flux
through 3-cycles of the orbifold space and non-trivial abelian
magnetic F-fluxes through 2-cycles supported on pairs of
D9-$\o{\rm D9}$ branes. Thus, we cancel the
localized tadpoles of both the O3- and the O7-planes in a
non-local way.
%As we will see the basic virtues of both approaches survive and do not
%mutually exclude each other:
At leading order, we will show that the F-flux through 2-cycles
can cancel all R-R tadpoles, with
chiral fermions and part of the K\"ahler moduli frozen via a
D-term potential. In addition, the G-fluxes through 3-cycles can
be chosen in such a way that additional complex structure moduli
are stabilized by an F-term potential. 
Since we are choosing the G-fluxes in a non-local
way, $\int H_3\wedge F_3\neq 0$, the Chern-Simons term in the
ten-dimensional effective IIB action will provide a G-flux
contribution to the Ramond charges, which has to be canceled by
the various non-local D-brane charges together with the negative
R-R charge of the orientifold planes.

Let us emphasize that in this paper we are only working in leading
order in string and sigma model perturbation theory. In
particular, since we are not cancelling the D7-brane charge
locally, we neglect the significant back-reaction on the
dilaton and on the background geometry, which, as we learned from
F-theory, is expected to lead to non-Ricci-flat manifolds.

The paper is organized as follows. In the next section we briefly
define our Type IIB $\ZZ_2\otimes \ZZ_2'$ orientifold model. Then,
in section 3, we  introduce the 3-form G-fluxes, compute their R-R
tadpoles, and discuss some general features of G-flux-induced
potentials. We stress the importance of mutual ``non-locality'' of
the R-R and NS-NS fluxes in avoiding ground states corresponding to
degenerate Calabi-Yau manifolds. In the next section, we work out
some details of the $\ZZ_2\otimes \ZZ_2'$ orientifold compactification
and discuss
the pattern of complex structure moduli stabilization and the
issue of supersymmetry breaking. Finally, in section 5, we
introduce D9-$\o{\rm D9}$ branes with F-fluxes and present 
brane configuration with chiral fermions, which satisfy  all
tadpole conditions. It should be mentioned that these  models are not
realistic; however it neatly demonstrates that models with
G-fluxes and D9-$\o{\rm D9}$ 
branes with F-fluxes (or intersecting D6-branes in
the T-dual picture) can be constructed in such a way that the
chiral fermions survive and both types of moduli can be at least
partially stabilized. We hope that our construction can serve as a
template for more realistic model-building.

\newsec{The Type IIB orientifold model}

In \rkst\ it was explicitly shown how turning on appropriate
3-form fluxes in a toroidal orientifold [$\Omega R(-1)^{F_L}$]
model can lead to supersymmetry breaking while freezing
some of the complex structure moduli, K\"ahler moduli,
and the dilaton. At the level of the four-dimensional effective
action, the freezing of the complex structure moduli and the
dilaton was due to the F-term potential \tv.  After turning on
this G-flux, the Chern-Simons term in the ten-dimensional Type IIB
effective action produces a tadpole for the R-R 4-form potential.
In general, additional D3-branes had to be present in order to
satisfy the R-R tadpole cancellation condition for the 4-form. The
resulting gauge theory on the D3-branes in the example discussed
in \rkst\ is always non-chiral. In view of applications to
realistic string model building, it is desirable to generalize such
flux compactifications to cases admitting chiral gauge theories.

It is known that a large class of chiral models is given by
intersecting brane world models or, in their T-dual version, by
D-branes with magnetic fluxes \refs{\bachas,\agnt}. Since we would
like to have the possibility to preserve supersymmetry, we are led
to flux compactifications on {\it orientifolds} of Calabi-Yau
threefolds. To be more precise, in this paper we consider a simple
orientifold model, namely we choose the Calabi-Yau to be given by
the orbifold \eqn\orbif{   X={T^6\over \ZZ_2\times \ZZ'_2}. } The
two $\ZZ_2$ operations are given by \eqn\acti{  \theta:\cases{
z^1\to -z^1 &\cr z^2\to -z^2 &\cr z^3\to
      z^3 &\cr} ,
            \quad\quad
            \theta':\cases{z^1\to z^1 &\cr z^2\to -z^2 &\cr z^3\to
                -z^3 &\cr}  }
on the three complex coordinates of $T^6=T^2\times T^2\times T^2$.
These data do not specify the orbifold completely, as we have the
freedom of introducing a discrete torsion, $\epsilon=\pm 1$
 in the $\ZZ_2$ twisted
sectors. Following \raaads\ we associate the model without
discrete torsion, $\epsilon=+1$, 
 to Hodge numbers $(h_{21},h_{11})=(51,3)$ and the
model with discrete torsion, $\epsilon=-1$,  to the mirror with
$(h_{21},h_{11})=(3,51)$. For $\epsilon=+1$, in the three
$\ZZ_2$ twisted sectors, the 3-cycles survive the projection, while
for $\epsilon=-1$, the 2-cycles survive. 
In the case with discrete torsion,
all 3-cycles on $X$ derive from 3-cycles on the ambient
$T^6$. The massless bosonic modes in the untwisted sector arise as
follows. From the ten-dimensional graviton one gets the
four-dimensional graviton, $g_{\mu\nu}$, and 9 scalars from the
internal components of the metric $g_{ab}$. Six of these scalars
form 3 chiral multiplets containing the 3 complex structure
moduli $\T^i$. The 3 remaining scalars combine with 3
additional scalars from the internal components of the R-R 4-form,
$(C_4)_{abcd}$, into 3 chiral multiplets, $C^I+i\cK_s^I$
containing the untwisted K\"ahler moduli $\cK_s^I$.  Finally, one also gets
the complex dilaton multiplet $C_0+i e^{-\phi}$. This untwisted
massless spectrum is enhanced by 48 chiral multiplets containing
the K\"ahler moduli related to the fixed points of the $\ZZ_2$
actions. In the case without discrete torsion, the chiral
multiplets from the twisted sectors are related to 48 additional
complex structure moduli.

In order to introduce objects that contribute negatively to the
R-R tadpole, we now take the additional quotient by $\Omega
R(-1)^{F_L}$, where $R$ reflects all three complex coordinates:
$z^I\to -z^I$. This breaks supersymmetry in the closed string
sector down to $N=1$  and introduces 64 O3-planes located at the
fixed points of $R$ in addition to three sets of 4 O7-planes
located at the fixed locus  of $R\theta$, $R\theta'$ and
$R\theta\theta'$.

There is a subtlety at this point, which deserves some clarifying
comments.  As shown in \raaads, the perturbative orientifold with
discrete torsion is forced to contain exotic orientifold planes in
order to satisfy the cross-cap constraint\foot{We thank
G. Pradisi for a discussion on this point.
We are indepted to A. Uranga for pointing out an error in an earlier
version of the paper.}
\eqn\con{     \int dl \langle \Omega R(-1)^{F_L}\, \theta |
e^{-l{\cal H}_{cl}} |
  \Omega R(-1)^{F_L}\, \theta' \rangle = \int {dt\over t}\, {\rm Tr}_{\theta\theta'}  \left
    (  \Omega R(-1)^{F_L}\, \theta \, e^{-2\pi t H} \right) .}
Such exotic orientifold planes have positive charge and tension.
More concretely, an odd number of the four classes of orientifold 
planes have to be of type Op$^{(+,+)}$.
%\foot{
%Since D-branes always have positive tension, 
%there is no way to cancel the R-R tadpoles in a supersymmetric way.}
Note that the intersecting brane world model discussed in
\refs{\rcvetica,\rcveticb} is T-dual to the Type IIB orientifold
model without discrete torsion.
% and therefore not to the model we
%will discuss here.

To proceed farther, we now allow turning on non-trivial 3-form
fluxes, which will contribute to the 4-form tadpole. In order
to cancel the three 8-form tadpoles, we have to introduce
additional D7-branes. Since we would like to discuss chirality, we
allow more generally the presence of D9-branes with magnetic
fluxes. 
%We can recover lower-dimensional branes in some infinite
%limit for the flux.

\newsec{Three-form fluxes}

In this section, we discuss the effect of turning on NS-NS and R-R
3-form fluxes on the internal Calabi-Yau manifold. We are
using the conventions and notation of \tv\ and \caf.

\subsec{R-R tadpole}

The Chern-Simons term in the Einstein-frame effective Type IIB
action looks like 
\eqn\csterm{   S_{CS}={1\over 2 \kappa_{10}^2}
\int_{M_4\times X} {C_4\wedge G \wedge
    \o{G}
   \over 4i\, \Im(\tau) }. }
%where the three form field strength is given in terms
%of the NS-NS threeform and the RR threeform as $G_3=\tau H_3 + F_3$.
Recall that $G=\tau\, H_3+F_3$, where  $H_3$ comes from the NS-NS
sector, $F_3$ from the R-R sector and $\tau=C_0 +ie^{-\phi}$ is
the dilaton which ``complexifies'' the flux.
The ten-dimensional gravitational coupling is given in terms of
$\alpha'$ as $\kappa_{10}^2={1\over 2}(2\pi)^7(\alpha')^4$. From \csterm\ 
it is
clear that turning on a non-trivial G-flux leads to a tadpole for
the 4-form $C_4$. In fact, the contribution to the tadpole is
given by \eqn\fluxcon{   N_{flux}={1\over 2\kappa_{10}^2\, \mu_3}
\int_X H_3\wedge F_3,}
with $\mu_p=(2\pi)^{-p}\, (\alpha')^{-{(p+1)/2}}$. To describe the
fluxes we assume that we have an integral basis of the homology
$H_3(X,\ZZ)$ with the intersections  $A^\Lambda\cap
A^\Sigma=B_\Lambda\cap B_\Sigma=0$ and  $A^\Lambda\cap
B_\Sigma=\delta^\Lambda_\Sigma$ ($\Lambda,\Sigma=0,\dots , h_{21}$). 
In terms of the Poincar\'e dual
basis of $H^3(X,\ZZ)$, $(\alpha_\Lambda, \beta^\Lambda)$, the
covariantly constant $(3,0)$ form can be expanded as \eqn\exom{
\Omega_{3}=X^\Lambda\, \alpha_\Lambda - F_\Lambda\, \beta^\Lambda,
} with \eqn\xxff{    X^\Lambda=\int_{A^\Lambda} \Omega_{3},
\quad\quad
              F_\Lambda=\int_{B_\Lambda} \Omega_{3} ,}
where the periods $X^\Lambda$ and $F_\Lambda$ are functions of the complex
structure moduli $\T^i$ ($i=1,\dots , h_{21}$).
Due to the Bianchi identities, the three-forms $H_3$ and $F_3$ are
closed, therefore in cohomology they can be expressed as integer linear
combinations 
\eqn\threeexr{\eqalign{     {\textstyle{{1\over
(2\pi)^2\alpha'}}}\, H_3&= e^{1}_\Lambda\, \beta^\Lambda +
    m^{1\Lambda}\, \alpha_\Lambda\cr
                 {\textstyle{{1\over (2\pi)^2\alpha'}}}\, 
     F_3&= e^{2}_\Lambda\, \beta^\Lambda +
    m^{2\Lambda}\, \alpha_\Lambda .\cr }}
Thus the  complex three-form flux can be written as
\eqn\hflux{  {\textstyle{{1\over (2\pi)^2\alpha'}}}\, G=
e_{\Lambda}\beta^{\Lambda}+ m^{\Lambda}\alpha_{\Lambda}}
with $e_{\Lambda}=\tau\, e^1_{\Lambda}+e^2_{\Lambda}$ and
      $m^{\Lambda}=\tau\, m^{1 \Lambda}+m^{2 \Lambda}$.
In this notation, the tadpole \fluxcon\ becomes \eqn\fluxiu{
N_{flux}= m\times e, } with $m\times e=m^{1\Lambda}\, e^{2}_\Lambda
- m^{2\Lambda}\, e^{1}_\Lambda$.

\subsec{The scalar potential}

The kinetic term for the G-flux
\eqn\kinterm{   S_G=-{1\over 4\kappa_{10}^2 \Im(\tau)} \int_X   G\wedge \star_6
  G,}
when integrated over the internal manifold, gives rise to a scalar
potential, which has been computed in \tv.
Note, that we use the convention that the Hodge star operator
involves also complex conjugation.
Working in the $(\alpha_\Lambda, \beta^\Lambda)$ basis the computation
is straightforward when one uses the following action of the Hodge star operator
\eqn\actiostar{\eqalign{   \star \alpha &= A\alpha + B \beta \cr
                            \star \beta &= C\alpha + D \beta ,\cr }}
where the four matrices $A,B,C,D$ can be expressed in terms of the
period matrix $\N$
\eqn\per{\eqalign{    A&=-D^T=(\Re\N)(\Im\N)^{-1} \cr
                      B&=-\Im\N-(\Re\N)(\Im\N)^{-1}(\Re\N) \cr
                      C&=(\Im\N)^{-1}. \cr }}
In a symplectic basis in which the prepotential $F$ exists \caf,
 \eqn\nmatrix{
\N_{\Lambda\Sigma}=\o{F}_{\Lambda\Sigma}+2i{
  {\Im(F_{\Lambda\Gamma})\, \Im(F_{\Sigma\Delta})\, X^\Gamma\, X^\Delta \over
            \Im(F_{\Gamma\Delta})\, X^\Gamma\, X^\Delta} } ,}
 where $F_{\Lambda\Sigma}=\partial^2 F /\partial X_\Lambda\,
\partial X_\Sigma $.

The scalar potential resulting from \kinterm\ can be expressed as
\eqn\vh{ V=-{\mu_3\over 2\,
\Im\tau} [m(\Im\N)\bar{m}+(e+m\Re\N)(\Im\N)^{-1}(\bar{e}
+\bar{m}\Re\N)]\, .}
Since the period matrix depends on complex structure moduli, $V$ is a 
function of $\T^i$ and $\tau$.
This scalar potential can also be rewritten as
\eqn\vlop{
V=-{\mu_3\over 2\, \Im\tau} [(e+m\bar{\N})(\Im\N)^{-1}(\bar{e}+\bar{m}
{\N})] +\mu_3\, \mte\, ,} or as
\eqn\vlom{
V=-{\mu_3\over 2\, \Im\tau} [(e+m\N)(\Im\N)^{-1}(\bar{e}+\bar{m}
\bar{\N})] -\mu_3 \, \mte\, .}
%where the constant ``cosmological'' term
%\eqn\mtedef{\mte\equiv m^{1\Lambda}e^2_{\Lambda}-
%m^{2\Lambda}e^1_{\Lambda}\, =(\Im\tau)^{-1}\,\Im (m\bar{e}).}
Recall that in our conventions, one requires $\Im\N<0$ in the
physical domain of positive-definite kinetic energy terms, while
$\Im\tau>0$.

In \tv, only the ``local'' case of $N_{flux}=\mte=0$ was
considered. Then it is easy to see that no stable minima of the
potential exist, except at some singular points (or limits) where
the period matrix degenerates. In order to understand the origin
of this result and the ``non-local'' case  of $\mte\neq 0$, we
first notice that there are two obvious candidates for the minima,
at \eqn\emnbar{e_{\Lambda}+m^{\Sigma}\bar{\N}_{\Sigma\Lambda}=0}
and at \eqn\emn{e_{\Lambda}+m^{\Sigma}\N_{\Sigma\Lambda}=0.} After
multiplying \emnbar\ by $\bar m^{\Lambda}$ and taking the
imaginary part, we obtain
\eqn\cond{m(\Im\N)\bar{m}=-\Im\tau\,(\mte),} hence Eq.\ \emnbar\ can
be satisfied in the physical positivity domain only if $\mte>0$.
By a similar argument, $\mte<0$ is a necessary condition for the
existence of a non-degenerate solution of Eq.\ \emn. Hence for a
given sign of $\mte$, only one of the two equations \emnbar\ and
\emn\ can be solved. Note that the potential is always positive at
the minimum point: $V_{min}=\mu_3 |\mte|=\mu_3 |N_{flux}|$.

In \tv\ is was shown that the first term in \vlop\ can be
understood as the F-term scalar potential arising from the
superpotential \refs{\rgvw,\rgukov }
\eqn\superpot{W={1\over \sqrt{2} \kappa_{10}} \, \int_X \Omega_3\wedge
G=\sqrt{\mu_3}\, \left(e_{\Lambda}X^{\Lambda}+
m^{\Lambda}F_{\Lambda}\right).}
In fact, by
using the identities \caf\ $F_{\Lambda}=\N_{\Lambda\Sigma}
X^{\Sigma}$, $D_iF_{\Lambda}=\bar{\N}_{\Lambda\Sigma}
D_iX^{\Sigma}$ and the tree-level K\"ahler potential, one can
rewrite \vlop\ as\foot{The $-3|W|^2$ term is canceled by K\"ahler derivatives
of the CY volume hypermultiplet.} 
\eqn\newsc{ V=\mu_3\, e^{[K(z,\o
z)+\tilde{K}(\tau,\o{\tau})]}\,
         \left[ G^{i\o{j}}\, D_i W D_{\o{j}} \o{W} +
         G^{\tau\o{\tau}}\, D_\tau W D_{\o{\tau}} \o{W}\right]+\mu_3\mte.}
Assuming that \emnbar\ is satisfied, \eqn\min{D_iW=D_{\tau}W=0;}
the existence of a supersymmetric minimum is therefore not guaranteed,
unless $W=0$ for some choice of fluxes.
%It has been shown that $\alpha'$ corrections to
%the K\"ahler potential as well as euclidean D3-brane space-time
%instanton corrections to the superpotential will eventually break
%the no-scale structure of the scalar potential
%\refs{\rbbhl,\rkklt}.
If the minimum is described by Eq.\ \emn\ instead of \emnbar, it
is convenient to replace \superpot\ by an ``almost holomorphic''
superpotential $\widetilde{W}=W(e{\to}\bar{e},m{\to}\bar{m})$, now
treating $\bar\tau$ as a chiral field. Such a superpotential
generates the potential \vlom, up to the $\mte$ constant
term.
It is easy to see that Eqs.\
\min\ are satisfied for this new superpotential, with
$\tau\to\bar\tau$.

We will see in section 5 that the topological $\mte$ term in
\newsc\ plays an essential r\^ole in the computation of  the
Fayet-Iliopolous terms for the abelian gauge groups
on  the D-branes, which have to be introduced to cancel
the R-R tadpoles.
The  $\mte$ term is proportional to the R-R three-form charge, which already
indicates that it is nothing else than the effective ``D3-brane''
tension of the G-flux.

Another, equivalent way of discussing supersymmetry breaking is by
examining the supersymmetry
transformations of fermions \rkst.
The condition for (at least one) unbroken supersymmetry can
be succinctly summarized as the requirement that $G$ be a pure $(2,\bar{1})$ [or
$(1,\o{2})$] form. Is this condition  satisfied in a vacuum
described by Eqs.\ \emnbar\ or \emn? It is easy to see
that Eq.\ \emnbar\ is equivalent to the condition that $G$ be
``imaginary anti-self
dual", i.e.\ $\star G=-i\bar G$;
%, where * denotes the Hodge star,\foot{Note that * acts
%on $\tau$ by complex conjugation.}
hence, in addition to the $(2,\o{1})$ part, it may also contain
$(0,\o{3})$; the latter may vanish, though, for a particular choice
of fluxes, which ensures $W=0$ at the minimum. Similarly, Eq.\ \emn\
describes an ``imaginary self-dual'' flux configuration, $\star
G=i\bar G$, a $(1,\bar{2})$ form with a possible $(3,\bar{0})$
admixture that vanishes if $\widetilde{W}=0$. Note that the
additional condition for supersymmetry, namely that the $G$ flux has
to be primitive, \eqn\primitiv{     J\wedge G =0, }
can be  satisfied on a
Calabi-Yau manifold, as there are no cohomologically non-trivial
closed 5-forms.

We conclude that a stable vacuum can only exist in the
``non-local'' case of  $\mte\neq 0$ i.e.\ with a non-vanishing
$N_{flux}$. This typically leads to a complete supersymmetry
breakdown; however, one (or more) supersymmetry may survive if the
moduli satisfy one additional constraint. Note that the number of
equations contained in \emnbar\ [as well as in \emn] is equal to
the number of undetermined moduli (including the dilaton), and that
moduli stabilization is therefore expected to occur for a generic
pattern of fluxes while unbroken supersymmetry would take place
only in some
special cases. In the next section we will describe examples
illustrating both kinds of situations.

\newsec{$\ZZ_2\otimes \ZZ_2'$ Orbifolds}

As an example we now apply the general formalism presented in section 2 to
our $\ZZ_2\otimes \ZZ_2'$ orientifold.

\subsec{Cohomological basis}

%We will consider type IIB compactifications on $T^6/(\ZZ_2\otimes \ZZ_2')$.
%The $T^6$ lattice consists of three planes, $(x^n,y^n)$, $n=1,2,3$.
%The orbifold group is defined by the rotations $g=e^{i\pi(J_1-J_2)}$
%and $g'=e^{i\pi(J_2-J_3)}$.
The following closed 3-forms on the toroidal ambient space are
invariant under the $\ZZ_2\times \ZZ_2'$ orbifold symmetry
\eqn\cohom{\eqalign{  & \alpha_0=dx^1\wedge dx^2\wedge
dx^3\qquad\quad \beta^0= dy^1\wedge dy^2\wedge dy^3\cr &
\alpha_1=dy^1\wedge dx^2\wedge dx^3\qquad\quad \beta^1=-
dx^1\wedge dy^2\wedge dy^3\cr & \alpha_2=dx^1\wedge dy^2\wedge
dx^3\qquad\quad \beta^2=- dy^1\wedge dx^2\wedge dy^3\cr &
\alpha_3=dx^1\wedge dx^2\wedge dy^3\qquad\quad \beta^3=-
dy^1\wedge dy^2\wedge dx^3.}} These are Poincar\'e-dual to the
obvious 3-cycles on $T^6$. Note that expanding $H_3$ and $F_3$
in terms of these eight 3-forms guarantees that the 3-form
fluxes in \threeexr\ are invariant under the orientifold symmetry
$\Omega\, R(-1)^{F_L}$.

There are three  moduli, $\T^i\equiv\R^i+i\I^i$, $i=1,2,3$, which
define the complex structure: $z^i=x^i+\T^i\,y^i$ (no summation
over $i$) on the orbifold space. As usual, the holomorphic
3-form \eqn\hol{\Omega_3=dz^1\wedge dz^2\wedge
dz^3=X^{\Lambda}\alpha_{\Lambda}- F_{\Lambda}\beta^{\Lambda}}
defines the homogeneous coordinates $X^{\Lambda}$ and the
derivatives $F_{\Lambda}=\partial_\Lambda F$ of the prepotential
\eqn\coord{\eqalign{   X^0&=1\phantom{\T}   \quad\quad
F_0=-\T^1\, \T^2\, T^3 \cr
                       X^1&=\T^1\quad\quad F_1= \T^2\, T^3 \cr
                       X^2&=\T^2\quad\quad F_2= \T^1\, T^3 \cr
                       X^3&=\T^3\quad\quad F_3= \T^1\, T^2. \cr }}
Therefore the prepotential is given by \eqn\prep{F={X^1X^2X^3\over
X^0}=\T^1\T^2\T^3.} It is convenient to introduce the following
basis of $(2,\bar{1})$ forms:
\eqn\cocom{\eqalign{&a_1=d\bar{z}^{\bar{1}}\wedge dz^2\wedge
dz^3,\cr &a_2=dz^1\wedge d\bar{z}^{\bar{2}}\wedge dz^3,\cr
&a_3=dz^1\wedge dz^2\wedge d\bar{z}^{\bar{3}}.}} By using the
prepotential \prep\ and Eq.\ \nmatrix, we obtain the following
(symmetric) period matrix: \eqn\ren{ \eqalign{ \N &= \;
\left(\matrix{ 2\R^1\R^2\R^3 & -\R^2\R^3& -\R^1\R^3& -\R^1\R^2\cr
\cdots&0& \R^3&\R^2\cr \cdots&\cdots&0&\R^1\cr
\cdots&\cdots&\cdots&0  \cr }\right)\cr & +i  \left(\matrix{
q\I^1\I^2\I^3
 & -\I^2\I^3{\R^1\over\I^1}& -\I^1\I^3{\R^2\over\I^2}& -\I^1\I^2{\R^3\over\I^3}\cr
 \cdots&{\I^2\I^3\over\I^1}&0&0\cr
\cdots&\cdots&{\I^1\I^3\over\I^2}   &0\cr
\cdots&\cdots&\cdots& {\I^1\I^2\over\I^3}  \cr }\right),}}
where \eqn\qdef{q=
1+\left({\R^1\over\I^1}\right)^2+\left({\R^2\over\I^2}\right)^2
+\left({\R^3\over\I^3}\right)^2.}
Note that $\I^i<0$ in the physical domain of $\Im\N<0$.

\subsec{Flux quantization on  orientifolds}

Flux quantization on orientifolds is quite subtle, since there exist
3-cycles ``smaller'' than the ones on the torus.
As defined above, the fluxes of $H_3$ and $F_3$ must belong to
$H^3(X,\ZZ)$, and respectively,
the homological Poincar\'e duals $[H_3]$ and $[F_3]$ must belong to $H_3(X,\ZZ)$.
\vskip 0.2cm
\noindent
{\it Fluxes for $\epsilon=+1$ orbifold}

Following the arguments used in \rbbkl, one notices that under the
$\ZZ_2\otimes \ZZ_2'$ and the $\Omega R (-1)^{F_L}$ action  a
toroidal 3-cycle $\pi_u$, in a general position, is mapped to an
orbit of eight toroidal 3-cycles all wrapping the same homology
class on $T^6$. Therefore, what is usually called a bulk 3-cycle
on the orientifold of $X$ is actually, from the toroidal point of
view, a cycle where  all $e_\Lambda$ and $m^\Lambda$ 
are multiples\foot{By that we actually mean $e_\Lambda^i$ and 
$m^{i\Lambda}$, $i=1,2$.} of 8. Hence the associated fluxes have
$N_{flux}$ quantized in 
multiples of 64.
Besides these bulk cycles totally inherited from
the ambient $T^6$, there are so-called twisted 3-cycles, which
also wrap some of the 48 3-cycles hidden in the various $\ZZ_2\times \ZZ_2'$
fixed points. Note that since the twisted cycles are ``shorter''
than the toroidal ones, the corresponding fluxes can carry smaller
quanta of $N_{flux}$. 
This is important for model-buliding since, as we will see later, the bulk
flux contribution, which is a quantum of $N_{flux}=64$, always exceeds the
negative contribution of the O3-planes to the R-R tadpole 
cancellation condition.
%To avoid the complications of twisted sector
%fluxes and moduli, in this paper we restrict ourselves
%to bulk 3-cycles. 
Hence supersymmetric configurations can only be obtained by switching on
some  twisted
fluxes (with $N_{flux}\le 32$).

\vskip 0.2cm
\noindent
{\it Fluxes for $\epsilon=-1$ orbifold}

For the model with
discrete torsion the situation is a bit more subtle.
In the following, we will argue that here $e_\Lambda$ and
$m^\Lambda$ are only quantized in units of 4. Again arguing
about the homology, $H_3(X,\ZZ)$,  we consider first
3-cycles on the $\ZZ_2\otimes \ZZ_2'$ orbifold space, i.e.
neglecting the orientifold projection for the moment. Naively, one
obtains that under the action  of $\ZZ_2\otimes \ZZ_2'$ a
bulk 3-cycle on the orbifold corresponds to an orbit of four toroidal
3-cycles, $4\pi_u$. One can show that the intersection numbers for
these  bulk cycles are multiples of 4. Since the twisted sectors
do not contain any additional 3-cycles, there must exist smaller
toroidal 3-cycles, as  an integral basis of $H_3(X,\ZZ)$
yields  a unimodular intersection form. Thus, we conclude that there
must exist 3-cycles in the orbifold, which correspond to only
toroidal orbits of length two: $2\pi_u$.

These shorter 3-cycles can be seen as follows.
Consider the $\theta$-twisted sector. In this sector, besides
the bulk 3-cycles,
we can define  fractional 3-cycles, which are of the form
$\pi_u + \pi_{tw}$,
where $\pi_u$ denotes a toroidal cycle and $\pi_{tw}$ a 3-cycle
in the $\theta$-twisted sector.
Under the action of the second $\ZZ_2$
such a fractional cycle is mapped to $\pi_u - \pi_{tw}$, so that the
whole orbit under $\ZZ_2\otimes \ZZ_2'$ gives rise to a pure
toroidal  cycle $2\pi_u$, which is indeed what we are looking for.

Taking also the $\Omega R (-1)^{F_L}$ action into account, we
conclude  that in the $\ZZ_2\otimes \ZZ_2'$
orientifold model with discrete torsion, the coefficients
in the expansion
\eqn\coeff{ {\textstyle{{1\over (2\pi)^2\alpha'}}}\, [G]=
e_{\Lambda}\, A^{\Lambda}+ m^{\Lambda}\, B_{\Lambda} }
are multiples of 4.
%This is the main reason why we consider in this paper mostly the
%model with discrete torsion, 
Therefore here the $N_{flux}$ is quantized
in multiples of 16, which does not exceed the contribution
from the O3-planes.

\subsec{Supersymmetry breaking}

In order to discuss supersymmetry breaking, we will examine the
transformations of fermions. The relevant terms depending on the
3-form background are \refs{\hw,\gp} \eqn\svar{\eqalign{  &
\delta\lambda\propto {\cal G}\epsilon+\dots\cr &
\delta\psi^m\propto\Gamma^m{\cal G}\bar{\epsilon}+2{\cal G}
\Gamma^m\bar{\epsilon}+\dots,}} where ${\cal G}\equiv
G_{abc}\Gamma^{abc},$ and $\epsilon$ is the supersymmetry
transformation parameter.\foot{We are using the notation of \hw.}
It is convenient to use the following form of $D=10$,
$32{\times}32$ gamma matrices: \eqn\gam{\Gamma^{\mu}= I\otimes
\gamma^{\mu}\qquad\qquad \Gamma^{i}= \gamma^i\otimes\gamma_5\, ,}
where $\gamma^{\mu}$ and $\gamma_5$ are $D=4$ gamma matrices. $I$
is the $8\times 8$ identity matrix; $\gamma^i$ are $D=6$
matrices for which we adopt the representation of \as:
\eqn\sen{\eqalign{ & \gamma^1=\half\sigma^1(1+\sigma^3)\otimes
{\bf 1}\otimes {\bf 1} \qquad\qquad\quad
\gamma^{\bar{1}}=\half\sigma^1(1-\sigma^3)\otimes {\bf 1}\otimes
{\bf 1}\cr & \gamma^2=-\sigma^3\otimes{i\over
2}\sigma^2(1+\sigma^3)\otimes {\bf 1}\qquad\qquad
\!\!\gamma^{\bar{2}}=-\sigma^3\otimes{i\over
2}\sigma^2(1-\sigma^3)\otimes {\bf 1}\cr &
\gamma^3=\sigma^3\otimes\sigma^3\otimes\half\sigma^1(1+\sigma^3)\qquad\qquad
\gamma^{\bar{3}}=\sigma^3\otimes\sigma^3\otimes\half\sigma^1(1-\sigma^3).}}
As in \as, we will denote the SO(2) spinor $\left(1\atop 0\right)$
by + and $\left(0\atop 1\right)$ by $-$. The $D=10$ spinor
$\epsilon$ is chiral, and can be written as
$\epsilon=\sum_{k=1}^{k=4}(\Psi^k\otimes\psi^k+
\bar{\Theta}^k\otimes\bar\theta^k)$, where $\Psi$ and $\bar\Theta$
are right-handed and left-handed SO(6) spinors, respectively,
i.e.\ containing  even or odd numbers of $(-)$, while $\psi$ and
$\bar\theta$ are similarly right- and left-handed in $D=4$.
However, the orbifold projection leaves us with SO(6) spinors
containing either only $(+)$ or only $(-)$, so in the absence of
fluxes, the compactified theory is $N=2$ supersymmetric, with
\eqn\sfour{\epsilon=(+++)\otimes\psi ~+~ (---)\otimes\bar\theta,}
%%~\equiv~\psi_{+++}+\bar{\theta}_{---}\, ,}
which can also be written as a sum of two Majorana-Weyl spinors
$\epsilon=\epsilon_L+i\epsilon_R$ with
\eqn\eps{\epsilon_A=(+++)\otimes\eta_A ~+~
(---)\otimes\bar\eta_A,} where $A=L,R,$ while $\eta_A$ are
right-handed components of $D=4$ Majorana spinors.\foot{Here, the
indices $A=L,R$ refer to the left- and right-moving parts of the
superstring and should not be confused with spacetime chirality.}

The $\Omega R (-1)^{F_L}$ orientifold projection leads to one more
restriction on the supersymmetry parameters \eps.
In particular it relates the left- and right-moving supersymmetries
because of
\eqn\leftright{   i\epsilon_R=i\Gamma^4\ldots\Gamma^9\, \epsilon_L .}
Thus, we see that the surviving spinor is 
%of the type considered in \gp\
\eqn\becker{  \epsilon=2 (---)\otimes\bar\eta_L .}

%which can be represented as a direct sum of two Majorana-Weyl
%spinors, \eqn\eps{\epsilon_A=(+++)\otimes\eta_A ~+~
%(---)\otimes\bar\eta_A,} where $A=1,2$ are the usual type IIB
%indices while $\eta_A$ are right-handed components of $D=4$
%Majorana spinors.%

%The $\Omega R (-1)^{F_L}$ orientifold projection leads to one more
%restriction on the supersymmetry parameters \eps. Acting on these
%spinors, $\Omega R (-1)^{F_L}=\Gamma=i\sigma_2\otimes
%I\otimes\gamma_5$, where $\sigma_2$ acts on the indices $A=1,2$.
%The orientifold projection dictates $(1-\Gamma)\epsilon=0$, therefore
%\eqn\constraint{\eta_2=i\eta_1,} and the surviving supersymmetry
%can be parameterized by a pure $(---)$ spinor,
%$\epsilon_1+i\epsilon_2$.

\subsec{An example with supersymmetry}

In the first example, we consider the following flux configuration:
\eqn\exa{\vec{m}^0=(0,4),\qquad\vec{e}_0=(-4,0),
\qquad\vec{m}^3=(-4,0),\qquad \vec{e}_3=(0,-4),}
or equivalently,
\eqn\exab{  {\textstyle{{1\over (2\pi)^2 \alpha'}}}\,  G=
 4(\alpha_0-\tau\beta^0-\tau\alpha_3-\beta^3)\, .}
%m^3=e_0=\tau,\qquad e_3=1,\qquad m^0=-1.}
Since $\mte=32$, we are looking for a solution of Eq.\ \emnbar, with
the period matrix given in \ren. Indeed, one can show that the
unique solution to these equations is \eqn\sola{\T^1\T^2=-1,
\qquad\qquad \tau=-\T^3.} The superpotential is \eqn\suppot{
W=4\sqrt{\mu_3}\, \left( -\tau-\T^3-\T^1\,\T^2\, \T^3 -\tau\, \T^1 \,
\T^2\right),} which vanishes for \sola. Moreover, all derivatives
satisfy \eqn\mina{D_iW=D_{\tau}W=W=0\, ,} so that the flux \exab\
is a pure $(2,\bar{1})$ form. This can explicitly be seen by
writing the G-flux as
\eqn\solb{  {\textstyle{{1\over (2\pi)^2 \alpha'}}}\, G={ 4{\T}^1\over
{\T}^1-{\o{\T}}^1}\, a_1+ {4\T^2\over \T^2-{\o{\T}}^2}a_2.}
Hence
we expect that $N=1$ supersymmetry remains unbroken.

In order to verify this, we examine the supersymmetry
transformations \svar\ with\foot{Note that, in our conventions,
$\bar{\epsilon}=(+++)\otimes\bar\psi ~+~ (---)\otimes\theta$.}
${\cal G}\propto\gamma^{\bar{1}}\gamma^2\gamma^3
+\gamma^{1}\gamma^{\bar 2}\gamma^3$. These variations are
restricted by
\eqn\van{\gamma^i (---)=\gamma^{\bar
\imath}(+++)=0.}
In fact, it is easy to see that all variations
under the $(---)$ transformations are zero, so that the
corresponding supersymmetry remains unbroken. Under the $(+++)$
transformations, the internal components $\psi^1$ and $\psi^2$ of
the gravitino have non-vanishing variations; therefore the
corresponding supersymmetry is broken. Note that the unbroken
supersymmetry is also preserved by the orientifold projection.
One can also check explicitly, that $G$ is indeed primitive.

Neglecting the orientifold projection,
the fluxes under consideration provide a nice example of partial
$N=2\to N=1$ supersymmetry breaking and moduli stabilization. This
partial breaking has  the same origin as in the APT mechanism
\apt\ in globally supersymmetric theories, where ``non-locality''
is due to the simultaneous presence of electric and magnetic
Fayet-Iliopoulos terms, which can be described by a
superpotential of the form \superpot.
Here, we have found a flux
configuration that preserves supersymmetry and fixes two combinations of
the four complex moduli $\{\tau,\T^1,\T^2,\T^3\}$. In the next
section, we will come back to this supersymmetry-preserving
example.

\subsec{An example with supersymmetry breaking}

As another example, we consider
\eqn\exc{\vec{m}^0=(0,4),\quad\vec{e}_0=(-4,0)\quad\Rightarrow\qquad
{\textstyle{{1\over (2\pi)^2 \alpha'}}}\, G=4(\alpha_0-\tau\beta^0).} 
Now \emnbar\ is solved by
$\R^1=\R^2=\R^3=\Re\tau=0$ and 
\eqn\exd{\Im\tau=-\I^1\I^2\I^3.} 
At this minimum, $D_iW=D_{\tau}W=0$; however, $W=-8\tau\neq 0$.
Furthermore, 
\eqn\exf{  {\textstyle{{1\over (2\pi)^2 \alpha'}}}\,  G={2\over
3}(a_1+a_2+a_3)+2\bar\Omega_3\, ,} 
and the flux therefore contains
also a $(0,\bar 3)$ contribution. As in the previous example, the
$(+++)$ supersymmetry is broken, but now
%with the goldstino supplied by the combination
%$\psi^1+\psi^2+\psi^3$ supplies the longitudinal degrees of freedom to the corresponding
%gravitino.
in addition, the $(---)$ supersymmetry is also broken. By looking
at the variations \svar, it is easy to check that it is indeed the
$(0,\bar 3)$ part of the flux that is responsible for this
breaking. From the point of view of the effective supergravity
theory, $N=1$ supersymmetry is broken by a vacuum expectation
value of the superpotential, i.e.\ of the auxiliary component of
the gravitational supermultiplet. This is a no-scale supersymmetry
breaking, at a scale undetermined at the classical level. A
detailed supergravity description of moduli stabilization and
partial supersymmetry breaking in similar orientifold models
has been worked out in \ferrara.
%The work of T.R.T. was supported in
%part by NSF grant PHY-99-01057.

\newsec{D-branes with magnetic fluxes}

So far, we have discussed
the consequences of turning on the background G-fluxes.
We have seen that they
contribute to the R-R tadpole cancellation conditions and that
a non-trivial superpotential is generated, which freezes some of the
complex structure moduli.
%This is in accord with the general structure
%that F-terms in such ${\cal N}=1$ Calabi-Yau models do only
%depend on the complex structure moduli, whereas the D-term
%depend on the K\"ahler moduli \refs{\rbdlr,\rkklma}.
In the papers on such flux compactifications
the R-R-tadpole conditions are trivially satisfied
by introducing D3- and D7-branes located on top of the orientifold planes.
In our case, this would lead to a consistent model, even though the
massless spectrum on the branes would be non-chiral.
If we have really phenomenological applications of such models
in mind, we have to introduce branes in such a way that
chiral fermions are generated.

A possible way of achieving this is by introducing D9-branes with
abelian magnetic fluxes \refs{\bachas,\rbgklnon,\raads}. Such
configurations are T-dual (mirror-symmetric) to intersecting
D6-branes, which have been discussed extensively in the recent
literature \refs{\rangles
\rbgklnon\raads\rbgklmag\ras\rafiru\rafiruph\rbkl\rimr\rbonna\rrab
\rott\rcvetica\rcveticb\BailinIE\rqsusy
\rqsusyb\rkokoa\rbbkl\rura\rbbklb\rpradisi\rbgo\CremadesQJ-\lustie}. 
Moreover, it is  known
that such abelian magnetic fluxes,
 via the integrated Dirac-Born-Infeld (DBI) action,
give rise to a D-term potential for the K\"ahler structure moduli $\cK_s^I$.
Therefore, one expects that  turning on both G-flux and magnetic
fluxes freezes both  some of the complex structure moduli $\T^i$ via
F-terms and K\"ahler structure moduli $\cK_s^I$ via D-terms.

\subsec{R-R tadpoles}

In order to cancel the 4-form and 8-form tadpoles arising from
the O3- and O7-planes in the $\ZZ_2\times \ZZ'_2$ orientifold
model, we introduce D9-branes with abelian magnetic fluxes.
First we notice that under the orientifold projection a D9-brane
with a constant magnetic flux, $F$, is mapped into a $\o{\rm D9}$ brane
with the opposite flux, $-F$. Therefore,  one might naively expect
that supersymmetry is broken by such branes. For pure D9-branes
this is indeed the case. However, by turning on constant
magnetic fluxes on the D9-branes, supersymmetric configurations
are possible.

The Chern-Simons term for  the D9-$\o{\rm D9}$ brane system
\eqn\csopen{   S_{CS}=\mu_9\, \int_{\rm D9}  {\rm Tr} \left( e^{2\pi \alpha' F}
\right) \wedge \sum C_q -\mu_9 \int_{\o{\rm D9}}  
{\rm Tr} \left( e^{-2\pi \alpha' F} \right)
\wedge \sum C_q  }
introduces possible tadpoles not only for the R-R 10-form but also
for lower-rank R-R forms. Here we have taken into account that
at the orbifold point (away from the singular fixed points)
the manifold is flat, so that the
curvature contributions to  the Chern-Simons term  vanish.
Note, that the R-R 10-form and the R-R 6-form tadpole cancels
automatically in \csopen.
To be  more explicit, we choose $K$ stacks of $N_a$ D9-branes,
$a=1,\ldots K$,
with the abelian magnetic fluxes
\eqn\fluxa{   {\F}_a=2\pi \alpha' F_a=2\pi \alpha'
\sum_{I=1}^3  F_a^I  dx^I\wedge dy^I  }
turned on on each brane. We use the normalization $\int_{T^2_I} dx^I\wedge dy^I=1$,
for each  $I=1,2,3$. Considering the T-dual situation with D6-branes at angles,
it is clear that we have two co-prime integers $(n^I,m^I)$ for each $T^2_I$ to specify
such a configuration
\eqn\speci{   F^I_a=2\pi{ n^I_a\over m^I_a} .}
The integer $m^I$ can be interpreted as the wrapping number
of the D9-brane around the 2-cycle $T^2_I$ and the second integer
$n^I$ is the first quantized Chern class of the $U(1)$ gauge bundle
\eqn\chern{   c_1(F_a^I)={1\over 2\pi} \int_{m_a^I\times T^2_I}
 F_a^I\, dx^I\wedge dy^I =n_a^I .}
Here we have taken into account that the D9-brane wraps
the two-dimensional torus $m^I$ times.
Under the action of $\Omega R(-1)^{F_L}$ such a brane with flux
is mapped to a $\o{\rm D9}$ brane with the opposite magnetic flux.
Moreover, we get lower-dimensional branes by choosing for the wrapping
numbers $m^I=0$.

So far we have only discussed the toroidal case. In the presence
of discrete torsion, the orbifold space contains also
48 additional 2-cycles from the three $\ZZ_2$ twisted sectors,
which in principle can also support non-vanishing  magnetic fluxes.
At the orbifold
point, such a brane would correspond to a fractional D9-brane,
which is also charged under twisted sector R-R fields.
To keep the presentation as simple as possible, we will
not consider such flux configurations in this paper and will
assume that the K\"ahler moduli related to these twisted
2-cycles are frozen at the orbifold point, i.e.\ the volumes
of the exceptional 2-cycles vanish.
Here we only consider magnetic 2-form fluxes through the
three 2-cycles inherited from the toroidal ambient space.
Arguments similar to the ones given for the allowed
3-cycles on the orbifold space lead to the conclusion
that such a bulk brane in the orbifold space can be described by
 four copies
of the  toroidal branes. This explains some extra factors of 4 in the
formulas presented below.

    From the Chern-Simons action, it is now straightforward to derive
the contribution of the D9-$\o{\rm D9}$  branes with fluxes to the
R-R tadpole cancellation condition. Taking also the contribution
from the orientifold planes and the G-flux into account, we arrive
at the following four conditions
\eqn\tadpolecm{\eqalign{ 8\sum_a N_a \prod_I n^I_a  + N_{flux} &= 32\cr
                     8\sum_a N_a n^1_a\, m^2_a\, m^3_a  &= -32\, \epsilon  \cr
                       8\sum_a N_a m^1_a\, n^2_a\, m^3_a  &= - 32 \cr
                       8\sum_a N_a m^1_a\, m^2_a\, n^3_a  &= - 32 \cr . }}
Here, the perturbative orientifold with discrete torsion ($\epsilon=-1$)
is assumed to contain one set of O7 planes of the type
O7$^{(+,+)}$. 

%As we
%mentioned  in section 2, the perturbative orientifold model has
%$N_{O3'}=64$, which can lead to brane supersymmetry breaking but
%spoils any chance of obtaining a supersymmetric brane
%configuration. Therefore,  we are choosing here $N_{O3'}=0$. 
The G-flux of course only contributes to the first line in \tadpolecm,
which is the R-R tadpole cancellation condition of the 4-form.
Moreover, the 10-form and 6-form tadpoles vanish automatically due
to the presence of the $\o{\rm D9}$ branes with opposite magnetic flux.
This is consistent with the fact that these forms are projected
out by $\Omega R (-1)^{F_L}$.

Using the Atiyah-Singer index theorem, or employing simply the results
of \rbgklnon, the chiral massless spectrum transforming
in the $U(N_1)\times \ldots\times U(N_K)$ gauge group is given in Table 1.
\vskip 0.8cm
\vbox{ \centerline{\vbox{ \hbox{\vbox{\offinterlineskip
\def\tablespace{height2pt&\omit&&
 \omit&\cr}
\def\tablerule{\tablespace\noalign{\hrule}\tablespace}

\hrule\halign{&\vrule#&\strut\hskip0.2cm\hfill #\hfill\hskip0.2cm\cr
& Representation  && Multiplicity &\cr
\tablerule
& $[{\bf A_a}]_{L}$  && ${1\over 2}\left(I_{a'a}+I_{Oa}\right)$   &\cr
\tablerule
& $[{\bf S_a}]_{L}$
     && ${1\over 2}\left(I_{a'a}-I_{Oa}\right)$ &\cr
\tablerule & $[{\bf ( N_a,\o N_b)}]_{L}$  && $4\prod_I \left(
m_a^I n_b^I - n_a^I m_b^I\right)$ &\cr \tablerule & $[{\bf (N_a,
N_b)}]_{L}$ && $4\prod_I \left( m_a^I n_b^I + n_a^I m_b^I\right)$
&\cr }\hrule}}}} \centerline{ \hbox{{\bf Table 1:}{\it ~~ Chiral
spectrum in $D=4$}}} } \vskip 0.5cm \noindent In Table 1 we used
\eqn\iise{\eqalign{   I_{a'a}&=32 \prod_I n^I_a m^I_a \cr
                      I_{Oa}&=8 \prod_I m^I_a -8\, \epsilon
                     \, m_a^1\, n_a^2\, n_a^3 -
                   8\, n_a^1\, m_a^2\, n_a^3-8\, n_a^1\, n_a^2\, m_a^3 .\cr}}
As we mentioned, extra factors of 4 appear because we only
consider bulk branes.

A comment on gauge anomalies is in order at this point.
For vanishing G-flux, the spectrum
in Table 1 is free of four-dimensional non-abelian gauge anomalies, if the R-R
tadpole
cancellation conditions are satisfied.
However, for non-trivial G-fluxes this is no longer true and the
chiral massless spectrum in Table 1 gives rise to the non-abelian
$SU(N_a)^3$ gauge anomaly:
\eqn\gaugeanom{ \delta_{\Lambda} \log Z_a=-{1\over 3!\, (2\pi)^2}\,
N_{flux}\prod_I m^I_a\, \int_{M^4}
     \left[ F_a\wedge F_a\wedge F_a \right]^{(1)} ,}
where we have used the Wess-Zumino descent relation notation, i.e.\
for a closed gauge invariant form $Y$, we define $Y=dY^{(0)}$ and
$\delta Y^{(0)}=dY^{(1)}$, where $\delta$ denotes a gauge variation.
As pointed out in \ruranga, this anomaly is canceled by an in-flow
mechanism resulting from the following term in the Chern-Simons action
of the D9-brane:
\eqn\cherns{    S_{CS}= \mu_9\int_{D9_a} C_2\wedge B\wedge {1\over 3!} \left(
       2\pi \alpha'\right)^3\, F_a\wedge F_a\wedge F_a  .}
By the usual descent relations, and taking the wrapping number of
the D9-brane into account, this leads to the anomalous gauge
variation \eqn\chernsb{    \delta_\Lambda S_{CS}={1\over 3!\,
(2\pi)^2}\, N_{flux}\prod_I m^I_a\, \int_{M^4}
     \left[F_a\wedge F_a\wedge F_a\right]^{(1)}, }
canceling precisely the naive anomaly \gaugeanom.

\subsec{Supersymmetry}

In order to identify the supersymmetry preserved by D9-branes with magnetic
fluxes, it is convenient to use the formalism of \MarinoAF.
Applying the results of \MarinoAF\ to one D9-brane with magnetic fluxes,
one finds that $N=1$ supersymmetry is preserved provided that
%some supersymmetry if the MMMS  equation \MarinoAF\  is satisfied
\eqn\susyt{\eqalign{&\sin (\varphi_a)\left({1\over 2} J\wedge J\wedge
\F_a  -{1\over 3!}
                       \F_a \wedge \F_a\wedge \F_a\right) +\cr
                       &\qquad\qquad\qquad
              \cos (\varphi_a)\left({1\over 2} J\wedge \F_a\wedge \F_a -
              {1\over 3!}
                       J \wedge J\wedge J\right) =0\cr} }
for any parameter $\varphi_a$. Here $J$ denotes the K\"ahler form
on the Calabi-Yau manifold. This can compactly be written as
\eqn\compact{   \Im\left(e^{-i\varphi_a} \Phi_a\right)=0, } with
\eqn\defphi{   \Phi_a={1\over 3!} (\F_a+iJ)\wedge (\F_a+iJ)\wedge
(\F_a+iJ) .} In this form the supersymmetry condition looks very
similar to the T-dual (mirror symmetric) condition for a 3-cycle
$\Gamma$  to be volume minimizing, $ \Im (e^{-i\varphi}
\Omega_3)|_\Gamma=0 $. The second  T-dual supersymmetric 
3-cycle condition, namely
that the 3-cycle is lagrangian, $J|_{\Gamma}=0$, is
automatically satisfied by flat 3-cycles that are mapped to our
D9-branes with constant magnetic fluxes.

Expanding the K\"ahler form in the string frame as \eqn\kaehler{
J=\sum_I  \cK_s^I \,  dx^I\wedge dy^I } and defining the angle
variables $\psi_a^I$  as 
\eqn\angles{ \tan\psi_a^I=2\pi\alpha'
{F_a^I\over \cK_s^I},} 
Eq.\ \susyt\ boils down to the familiar
supersymmetry condition 
\eqn\famsusy{  \sum_I \psi_a^I ={3\pi\over
2} -\varphi_a \quad {\rm mod}\ 2\pi.} 
As is well known,  changing one of the angles $\psi_a^I$ and
consequently also $\varphi_a$ by $\pi$ turns
the D9-brane into a $\o{\rm D9}$ brane. 
%Therefore, in order to describe really D9-branes and not 
%$\o{\rm D9}$-branes with flux
%we restrict the winding numbers to the range $n_a^I\ge 0$ and
%correspondingly the angles, $\varphi_a$, to the interval
%$-{\pi\over 2}<\varphi_a\le {\pi\over 2}$. 

Then, the  supersymmetry preserved by a  brane satisfying
\famsusy\ is given by 
\eqn\susybrane{  \epsilon_R=\Gamma_0 \ldots \Gamma_9
\,\rho(M)\, \epsilon_L ,} 
where $\rho(M)$ denotes the rotation
matrix by the angles $\psi_a^I$ in the spinor representation
\refs{\rpark,\rwitti}. 
Thus, a D9- brane preserves the same supersymmetry as the
orientifold plane \leftright, if 
\eqn\susybraneb{  \epsilon_L= i
\gamma_5 \rho(M)\, \epsilon_L.} 
Inserting the form of the
left-moving spinor  \eps, we arrive at the eigenvalue equation
\eqn\susyfinal{   \rho(M)\, \eta_L=-i \eta_L ,} 
which has a
solution if 
\eqn\soli{ \sum_I \psi^I_a=3\pi/2 \quad {\rm mod}\ 2\pi.} 
Therefore, the
$N=1$ supersymmetry preserved by the orientifold planes
corresponds to $\varphi_o=0$ and therefore, in order to preserve
the same supersymmetry, we have to choose  $\varphi_a=0$ for all
$K$ stacks of D9-$\o{\rm D9}$ branes.

\subsec{The scalar potential}

The DBI action for a stack of D9-branes with magnetic flux,
when integrated over the internal volume, gives rise to a scalar
potential of the form 
\eqn\dbiexact{ {\cal V}_a=\mu_9\, e^{-\phi}\, N_a\, 
\int_X  d^6 x \sqrt{
      {\rm det}[G+2\pi\alpha'F]} ,}
which for constant abelian fluxes is exact to all orders in
$\alpha'$. This potential holds for both supersymmetric and
non-supersymmetric  D9-branes but
for supersymmetric gauge fluxes satisfying \susyt\ the
computation simplifies, as the
square-root in \dbiexact\ simplifies.
In analogy to the T-dual situation for calibrated
3-cycles, for a supersymmetric D9-brane, this action can be
expressed as 
\eqn\dbisc{   {\cal V}_a=\mu_9\, e^{-\phi}\, N_a\,\int_X
\Re\left(e^{-i\varphi_a}
    \Phi_a\right) .}
Since we have computed the scalar potential in the closed string
sector in the Einstein frame we should also shift to the Einstein
frame in the open string sector. Defining the Einstein frame
K\"ahler moduli as $\cK_E=e^{-\phi/2}\, \cK_s,$ we can bring  the
scalar potential for  $K$ stacks of supersymmetric D9-$\o{\rm D9}$
branes with angles $\varphi_a$ into the suggestive form
\eqn\totpotnine{\eqalign{ {\cal V}_{D9}=& \,8 \sum_a {\cal V}_a \cr
=& \,8\,\mu_3\,   \sum_a
\cos\varphi_a\, N_a \prod_I n^I_a
                      -8\,\mu_7\,  e^{\phi} \sum_a
              \cos\varphi_a\, N_a \sum_{I\ne J\ne K} n^I_a\, m^J_a\, m^K_a\,
                       \cK_E^J\, \cK_E^K \cr
        &-8\,\mu_9\, e^{{3\over 2}\phi}  \sum_a \sin\varphi_a\,
       N_a \prod_I m^I_a  \cK_E^I\,
                      +8\,\mu_5\,  e^{{1\over 2}\phi} \sum_a
              \sin\varphi_a\, N_a \sum_{I\ne J\ne K} n^I_a\, n^J_a\, m^K_a\,
                       \cK_E^K .\cr}}
Thus we see that for generic angles $\varphi_a$, one gets
effective D9-, D7-, D5-  and D3-brane tensions. Note that the
extra factors of $(2\pi\alpha')$ in the magnetic flux $\F$ and the
powers of the dilaton arrange themselves in just the right way to
give the effective Einstein frame Dp-brane tensions. 

Moreover, \totpotnine\ is correct for both D9- and $\o{\rm D9}$-branes
by defining $\varphi_a$ appropriately. 
Starting for instance with a supersymmetric D9-branes with quantum numbers
$(n_a^I,m_a^I)$ and angle $\varphi_a$, reflecting an odd number of such pairs
gives the anti-brane for which we must choose $-\varphi_a$ in
\totpotnine. 

Adding also
the contributions from the O3- and O7-planes
\eqn\totpoto{\eqalign{ {\cal V}^E_{O}=-32\mu_3  - 32\,\mu_7\, e^{\phi}
                   \left( \epsilon\, \cK_E^2\, \cK_E^3 +
                          \cK_E^1\, \cK_E^2+\cK_E^1\, \cK_E^3 \right),  }}
and the topological term from the flux-induced potential\foot{Here we
assume that $N_{flux}$ is positive; all equations can be consistently
rewritten for negative $N_{flux}$.}
\eqn\fluxindu{  {\cal V}_{flux}=\mu_3 N_{flux},} and, in addition,
invoking the R-R tadpole cancellation conditions \tadpolecm, we
can write the combined potential 
${\cal V}_{D9}+{\cal V}^E_O+{\cal V}_{flux}$
as 
\eqn\sclacompl{ \eqalign{
{\cal V}_{FI}=&8\,\mu_3\,   \sum_a (\cos\varphi_a-1) \, N_a
\prod_I n^I_a
                      -8\,\mu_7\,  e^{\phi} \sum_a
           (\cos\varphi_a-1)\, N_a \sum_{I\ne J\ne K} n^I_a\, m^J_a\, m^K_a\,
                       \cK_E^J\, \cK_E^K \cr
        &-8\,\mu_9\, e^{{3\over 2}\phi}  \sum_a \sin\varphi_a\,
       N_a \prod_I m^I_a\, \cK_E^I
                      +8\,\mu_5\,  e^{{1\over 2}\phi} \sum_a
              \sin\varphi_a\, N_a \sum_{I\ne J\ne K} n^I_a\, n^J_a\, m^K_a\,
                       \cK_E^K .\cr}}
As was discussed in \rqsusy, in the effective four-dimensional
theory, such a potential originates from the Fayet-Iliopolous
terms associated to the $K$ abelian $U(1)\subset U(N_a)$
subgroups. By writing the potential as 
\eqn\fayet{   {\cal
V}_{FI}=\sum_a {N_a\over 2 g_a^2}\, \xi_a^2 } 
one can identify the
Fayet-Iliopolous terms $\xi_a$. 
%They vanish precisely when all
%D9-branes preserve the same supersymmetry as the orientifold
%planes, i.e.\ $\varphi_a=0$ for all stacks of branes. 
This
computation shows that the topological term  from the flux-induced
potential is crucial indeed and participates in the cancellation
of Fayet-Iliopoulos terms in the D-term potential.
%As it is well established in the context of
%intersecting brane world models, changing the K\"ahler structure
%such that supersymmetry is broken corresponds in the effective
%theory to a Fayet-Iliopolous term for the abelian gauge field
%inside $U(N_a)$. Therefore, the scalar potential induced by
%abelian magnetic fluxes corresponds to a D-term potential.
%Again this is in agreement with the general
%expectation that the D-term potential only depends on the
%K\"ahler moduli.

The positive-definite D-term potential vanishes 
precisely when the magnetic
flux preserves supersymmetry, which means that $\Im(\Phi_a)=0$ for
all $K$ stacks of D9-branes.
%\foot{Of course such a globally
%supersymmetric D-brane configuration is only possible for the
%choice $\epsilon=+1$ with G-fluxes $N_{flux}\le 32$.
%Remember that, due to the flux quantization conditions, 
%for $\epsilon=+1$ the bulk fluxes always satisfy 
%$N_{flux}>32$, so that supersymmetric models with non-trivial 
%G-flux can only be realized by  turning on twisted G-flux}. 
This leads to the following constraint for the string frame K\"ahler moduli: 
\eqn\kaehlersus{
-\prod_I \cK_s^I\, m_a^I +(4\pi^2\alpha')^2\, \sum_{I\ne J\ne K}
\cK_s^I
  m_a^I\, n_a^J\, n_a^K =0  }
for all stacks of D9-branes. In general, one has more conditions than K\"ahler
moduli, so that this condition drastically 
constrains the model\foot{Note that actually in the D-term potential
VEVs for charged fields can in principle cancel the Fayet-Iliopolous
term and lead to new minima of the scalar potential \agnt.
However, this means that D-branes recombine and that the gauge and matter
sector of the theory changes. For this not to happen we set
all charged open string moduli in
the D-term to zero by hand. Then supersymmetry gives rise to the condition
\kaehlersus}.

One might wonder whether the gauge fluxes also induce a
superpotential of the form \eqn\ftermgauge{    W=\int
\Omega_{3}\wedge \omega_{YM} ,} where $\omega_{YM}$ denotes the
Chern-Simons form \eqn\omeym{  \omega_{YM}=\sum_a A_a\wedge F_a .}
Here, it should be recalled that  \ftermgauge\ arises from the
general superpotential $W=\int \Omega_{3}\wedge G$ after taking
into account that, in general, gauge fluxes on D9-branes induce a
source for G-fluxes via the equation \eqn\sourci{ dG\sim \sum_a
F_a\wedge F_a .} In fact, if the right-hand side happened to be
non-zero, this would invalidate our whole reasoning, as $G$ would
have to be combined with $\omega_{YM}$ to give a closed 3-form (as
in Type I or in the heterotic string). However, in our orientifold
example, the contributions to the right-hand side of \sourci\ from
the D9-branes and  their image $\o{\rm D9}$ branes just cancel, so that
the magnetic fluxes only induce a D-term potential.

\subsec{Examples}

%Utilizing the G-flux from the section 4.4, we construct
%two simple examples in the following.
\vskip 0.2cm
\noindent
{\it 5.4.1 A supersymmetric brane configuration}

Here we discuss a 
supersymmetric brane configuration. 
Choosing the orbifold without discrete torsion, $\epsilon=+1$,
 and switching on a
G-flux\foot{As explained before, such
a configuration requires twisted 3-cycles. We assume that the complex stucture
moduli can be stabilized in a supersymmetric vacuum, like in the example
discussed in Section 4.4.} 
with $N_{flux}=32$
we introduce 
two stacks of four  D9-branes with
 the quantum numbers
\eqn\wrap{\eqalign{   {1^{st}~\rm stack~}&: (n^I,m^I)=\{ (0,1),
(1,-1),(1,-1)\} \cr
                      {2^{nd}~\rm stack~}&: (n^I,m^I)=\{ (1,0), (0,1),(0,-1)\} .\cr }}
One realizes that all four R-R tadpole cancellation conditions are
satisfied. The chiral spectrum on the D9-$\o{\rm D9}$ branes consists of four
chiral multiplets in the $({\bf \o{4}},{\bf 4})$ representation and another
four chiral multiplets in the $({\bf \o{4}},{\bf \o{4}})$ representation of the
$U(4)\times U(4)$ gauge group. Note that the second  stack of
branes has $m^{(1)}=0$ on the first torus and therefore can be
considered as   D7-branes localized on the first $T^2$. Moreover,
as expected from the general anomaly formula \gaugeanom, only the
first $SU(N_a)$ gauge factor is anomalous.

The second  stack of branes is supersymmetric for any choice of the 
K\"ahler moduli,
while the first stack yields
the constraint
\eqn\secondsusy{     \cK_s^{(2)}\, \cK_s^{(3)}=(4\pi^2\alpha')^2.}
Thus, the number of unfrozen K\"ahler parameters from the untwisted sector
 is reduced to two. We expect that one combination of the 4-form
 superpartners, $C^I$,
also gets a mass via axionic couplings from the Chern-Simons
action \refs{\rafiru,\rqsusy}. Note, that in these examples the
overall volume of $T^6$ is not frozen.

\vskip 0.2cm
\noindent
{\it 5.4.2 A non-supersymmetric brane configuration}

Consider the supersymmetric G-flux discussed in Section 4.4.
Since $N_{flux}=32$, the four-form tadpole cancellation condition
is already saturated without any D3-brane charges.
We are dealing now with the model with discrete torsion, $\epsilon=-1$,
which contains the O$7^{(+,+)}$ planes.
We introduce the following two stacks of four D9-branes with
 the quantum numbers
\eqn\wrap{\eqalign{   {1^{st}~\rm stack~}&: (n^I,m^I)=\{ (0,1),
(1,-1),(1,-1)\} \cr
             {2^{nd}~\rm stack~}&: (n^I,m^I)=\{ (1,0), (0,-1),(0,-1)\} .\cr }}
Indeed the R-R tadpole cancellation conditions are satisfied,
where the second stack can be seen as $\o{\rm D7}$-branes, which have
been introduced to cancel the R-R charge of the O7$^{(+,+)}$ plane. 

The gauge group is $U(4)\times U(4)$ and 
the chiral spectrum on the D9-$\o{\rm D9}$ branes consists of four
chiral fermions  in the $({\bf 4},\o{\bf 4})$ representation,
four chiral fermions in the $({\bf 4},{\bf 4})$ representation and eight
chiral fermions in the $({\bf 6},{\bf 1})+({\bf \o{10}},{\bf 1})$ 
representation.

We can make the FI-term of the first $U(1)\subset U(4)$ vanish
by satisfying  
the constraint
\eqn\secondsusynn{     \cK_s^{(2)}\, \cK_s^{(3)}=(4\pi^2\alpha')^2.}
However, in contrast to the previous example, 
the scalar potential (FI-term) for the second gauge factor does not vanish, 
but takes the value
\eqn\fayetsec{   {\cal V}_{FI,2}=64\mu_7\, \cK_s^{(2)}\, \cK_s^{(3)} =
                                 64\mu_3.}
It is beyond our scope to discuss this potential in more detail; it is 
clear that non-trivial magnetic fluxes can fix part
of the K\"ahler moduli.

\vskip 0.5cm

These simple examples show that a combination of G-flux and
magnetic fluxes on D9-$\o{\rm D9}$ branes allows  some of the conceptual
constraints for phenomenologically interesting string models to be
satisfied. On the one hand, similar to what takes place in
intersecting brane worlds, one can obtain a chiral spectrum on the
branes. On the other hand, the G-flux can fix some or maybe even
all complex structure moduli via an F-term potential, and the
abelian magnetic fluxes freeze some or even all of the K\"ahler
moduli via an effective D-term potential.
% If the D-term potential
%freezes all K\"ahler moduli, generically one expects them to be of
%the order of the string scale, which would spoil in general the
%supergravity approximation for the F-term scalar potential.
Furthermore, the remaining $N=1$ supersymmetry can be
spontaneously broken in the large radius limit  in a no-scale
manner, provided that one chooses a suitable G-flux configuration,
like in the example described in Section 4.5.

Of course, for bulk branes we obtain, in addition, open string
moduli related to continuous Wilson lines on the D9-branes along
the internal directions. These are not frozen in the leading-order
approximation we are discussing here. One way to freeze them
already at leading order would be to also turn on magnetic fluxes
through the $\ZZ_2\times \ZZ_2'$ fixed points. These would give
rise to fractional D9-branes with frozen open string moduli. It
might be that the back-reaction on the geometry freezes (some
of) these moduli too.

\newsec{Conclusions}

In this work, we have pointed out that by combining ideas from
Calabi-Yau compactifications with background fluxes and from
intersecting brane-world models, it is possible to freeze both
complex structure moduli and K\"ahler moduli. The resulting
gauge/supergravity theory has many phenomenologically appealing
properties: chiral spectrum, partial supersymmetry breaking and,
eventually, complete supersymmetry breaking in a no-scale manner,
with a vanishing cosmological constant. We have illustrated these
issues on a simple example provided by the $\ZZ_2\times \ZZ_2'$
toroidal orbifold. Of course, these concepts are more general and
can in principle be applied to any Calabi-Yau orientifold model
with O3-planes.

A very important point is that our computations are valid only
in the leading order of string perturbation theory.
Since the superpotential depends only
on the complex structure moduli, it is exact to all orders
in $\alpha'$ (although $\alpha'$ corrections to the  K\"ahler potential
will eventually also change the F-term scalar potential).
In addition, our computation for the D9-branes
was exact to all orders in $\alpha'$, as we used the complete
DBI action.

Taking the back-reaction of the G-flux into account, it is known
that it leads to a warping of the Calabi-Yau geometry,  while
leaving the dilaton constant. This seems to be promising; however,
the D9-branes with flux and the O7-planes do not cancel their R-R
charges and tensions locally (which was essential for generating
chiral asymmetry) so that their back-reaction will dramatically
change the background geometry and will make the dilaton vary over
the internal space. 
Thus, at this level, we cannot trust any longer the computation done 
in this work.
As we know from F-theory, the background is presumably
no longer Calabi-Yau; nevertheless, the hope is that
the string background adjusts itself in such a way
that the number of stabilized moduli remains determined
by the leading order terms, and that the gauge  group and the
chiral matter content are not changed. It would be interesting to understand
in more detail what the back-reaction is and what other moduli
might be frozen by it. 

Eventually,  it may be
worthwhile to establish duality relations to other types of flux
compactifications, in particular to
Type IIA orientifolds, to M-theory
or to the heterotic superstring.

%Thus, at this state we are losing control over
%the model. As we know from F-theory, the background is presumably
%no longer Calabi-Yau, but the hope is that, after all, the string
%background adjusts itself in such a way that indeed the number of
%moduli is at least reduced in the way computed from the
%leading-order terms, and that the gauge  group and the chiral
%matter content are not changed. It would be interesting to
%understand in more detail what the back-reaction is and in which
%way it changes the picture presented in this paper. 
%\vskip 5mm
\vfill\eject
\centerline{\bf Acknowledgements}
\vskip 3mm\noindent 
We are
grateful to Carlo Angelantonj, Ignatios Antoniadis, Sergio
Ferrara, Gianfranco Pradisi, Stephan Stieberger, Angel M. Uranga,
Pierre Vanhove and Anders Westerberg for very useful discussions.
T.R.T.\ is grateful to the Institute of Physics at Humboldt
University in Berlin, where this work was initiated,  for its kind
hospitality. His research is supported in part by the National
Science Foundation under grant PHY-99-01057. 
The work of R.B. and D.L. is supported in part by the EC under the RTN project
HPRN-CT-2000-00131.
\vfill\eject

\listrefs

\bye
\end